\documentclass[%
 aip,
 amsmath,amssymb,
 reprint,%
]{revtex4-1}
\usepackage{multirow}
\usepackage{tikz}
\usepackage{graphicx}
\usepackage{dcolumn}
\usepackage{bm}
\usepackage{comment}

\usepackage[utf8]{inputenc}
\usepackage[T1]{fontenc}
\usepackage{mathptmx}
\usepackage{etoolbox}
\usepackage{caption}
\usepackage{subcaption}
\usepackage{fancyhdr}
\usepackage{blindtext}
\usepackage{float}
\usepackage{tikz}
\usepackage{xcolor} 

\newcommand*{\coline}[1]{
\noindent
\begin{tikzpicture}
\draw[very thick, color=#1] (0,0) -- (0.3,0);
\end{tikzpicture}
}

\newcommand*{\cudot}[1]{%
    \begin{tikzpicture}
    \fill[#1] circle (2pt); 
    \end{tikzpicture}%
}

\definecolor{1}{rgb}{1.0000,0.6250,0} 
\definecolor{2}{rgb}{1.0000,0.3021,0}    
\definecolor{3}{rgb}{ 0.9792,0,0}   
\definecolor{4}{rgb}{0.6562,0,0}  
\definecolor{5}{rgb}{0.3333,0,0}  
\definecolor{6}{rgb}{0.0104,0,0}  
\definecolor{7}{rgb}{0.00,0.00,1.00}  
\definecolor{8}{rgb}{0,0,0}  

\makeatletter
\def\@email#1#2{%
 \endgroup
 \patchcmd{\titleblock@produce}
  {\frontmatter@RRAPformat}
  {\frontmatter@RRAPformat{\produce@RRAP{*#1\href{mailto:#2}{#2}}}\frontmatter@RRAPformat}
  {}{}
}%
\makeatother
\begin{document}

\preprint{AIP/123-QED}

\title[Starting vortex strength in an impulsively started airfoil]{Starting vortex strength in an impulsively started airfoil}
\author{A. Goyal}
 \affiliation{Department of Mechanical Engineering, McGill University, Montreal, Canada, H3A 0G4}
\author{J. Nedi\'{c}}%
 \email{jovan.nedic@mcgill.ca}
 \affiliation{Department of Mechanical Engineering, McGill University, Montreal, Canada, H3A 0G4}

\date{\today}

\begin{abstract}
This work characterises the starting vortex and unsteady shear layer generated by an impulsively started National Advisory Committee for Aeronautics 0010 (NACA 0010) airfoil at angles of attack ranging from $3^{\circ}- 7^{\circ}$. Measurements are obtained by using time resolved Particle Image Velocimetry. The net circulation in the field is accurately predicted by Wagner's function. It is observed that the starting vortex detaches from the shear layer whilst the airfoil is surging. Starting vortex circulation at detachment is presented for a range of surge speeds at pre-stall angles of attack for a given surge distance. Kinematic conditions resulting in starting vortex detachment are examined qualitatively through Finite Time Lyapunov Exponents and quantitatively through a non-dimensional velocity ratio. 
Furthermore, the formation of secondary Kelvin-Helmholtz-type vortices in the shear layer is found to depend on the Reynolds number rather than the detachment of the starting vortex.
\end{abstract}

\maketitle
\section{\label{sec:level1}Introduction:\protect\\ }
Vortices are ubiquitous in several natural and engineering systems. For example, jellyfish employ vortex rings for propulsion~\cite{dabirijellyfish}, insects rely on the leading edge vortex to generate lift~\cite{eldredge2019leading} and tip vortices generated at the aircraft wing tip contribute to overall drag. The relative motion between a body and a fluid leads to the formation of a shear layer, which, in turn, rolls up to form a drag vortex. Vortex strength, or circulation, dictates forces and moments acting on the body. Large vortex circulation is desirable in some scenarios, such as prolonging dandelion drift~\cite{dandy}, enhancing the lift generated by insect wings~\cite{rival2010influence}, and undesirable in some systems, such as gust encounter in micro aerial vehicles~\cite{gust}, as well as tidal stream turbines~\cite{tidal_turbines}. A better understanding of the vortex formation process enables more accurate reduced-order models that incorporate the most dominant flow features, improving the systems' engineering design. 

During formation, the primary vortex grows in size and circulation by entraining vorticity provided by a feeding shear layer. The growth is not indefinite and there is an upper limit to vortex circulation for a given set of initial conditions~\cite{gharib1998universal}. In a vortex generator which employs a cylinder-piston apparatus to push fluid through a nozzle, mean piston speed, $\overline{U}_{p}$ and nozzle diameter, $D$, are used to non-dimensionalise time. Non-dimensional time, $t^{*}=t\times \overline{U}p/D$ simplifies to $t^{*}= L/D$, where $L$ is the length of the ejected fluid column obtained from $\int_{0}^{t}u_{p}(t)\textrm{d}t$. ~\citet{didden1979formation} had originally devised this model, also known as the slug flow model, in an attempt to link vortex ring circulation to the initial conditions of the vortex generator, namely the ratio of the length of the ejected fluid column to a natural length scale of the vortex generator. 

The formation number is the non-dimensional time corresponding to an upper limit on primary vortex circulation. In several studies, the formation number was found to be around 4~\cite{gharib1998universal, jeon2004relationship, ringuette2007role, milano_gharib, dabiri2005starting}. ~\citet{gharib1998universal} had asserted that the value of 4 was universal. The concept of a universal formation number was further supported by the Kelvin-Benjamin variational principle which says that for a given impulse and vorticity distribution, there exists a stable energy maximum, beyond which the vortex generator cannot supply the energy required by the primary vortex ring~\cite{thomson1883treatise, benjamin1976alliance}. At this time, termed pinch-off, the primary vortex ring detaches from its feeding shear layer, thereby completing the formation process. The Kelvin- Benjamin principle provides a dynamic argument to describe pinch-off, \textit{i.e.,} the attainment of a local maximum in energy. Such a description obscures the critical kinematic condition that must be met for pinch-off to occur. ~\citet{shusser2000energy} provided the kinematic argument that pinch-off takes place because the convective speed of the primary vortex ring exceeds the velocity of its feeding shear layer. This kinematic argument also holds for orifice-generated vortex rings, where the feeding shear layer breaks into secondary vortices \cite{limbourg2021asymptotic}. Several experiments following ~\citet{gharib1998universal}'s work tested the universality of the formation number for different vortex generators and found different values of the formation number. ~\citet{devorianringuette} studied vortex formation due to a rotating low aspect ratio flat plate fin and reported a formation number of 0.2-0.7. ~\citet{dabiri2005starting} found values between 5-8 for vortex rings generated by variable nozzle diameters. For orifice-generated vortex rings, ~\citet{limbourg2021formation} observed a formation number of 2. The discrepancy in the formation number can be attributed to the fact that ~\citet{gharib1998universal} had linked the final state of the fully formed vortex ring to the initial slug flow conditions, without considering the intermediate formation process which strongly depends on the vortex generator. \citet{dabiri2009optimal} subsequently developed a formation number which included a factor to account for the vortex generator \textit{a posteriori}, $C\propto (\textrm{d}\Gamma/\textrm{d}t)^{-1}$. \citet{limbourg2021extension} later presented a unifying formation number for orifice and nozzle geometries, which directly incorporated the geometric properties \textit{a priori} via the \textit{vena contracta} effect on the starting flow. This model was then successfully used to predict the flow invariants for a variety of outlet geometries \cite{limbourg2021extended}. 
\begin{figure*}[t!]
\includegraphics{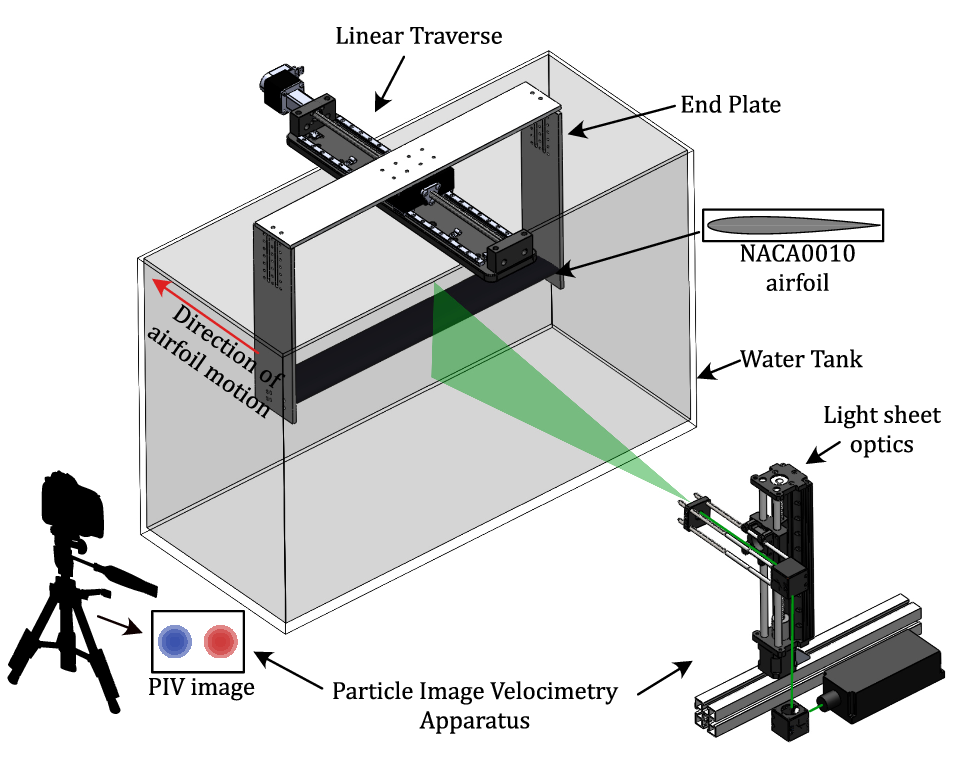}
\caption{Experimental setup for the present study, consisting of a NACA0010 airfoil, laser sheet generating apparatus and a high-speed camera.}
\label{fig:setup}
\end{figure*}

The majority of prior studies on vortex formation have been from axisymmetric geometries or those where there is some form of rotational symmetry e.g., \citet{o2014pinch}. Starting vortices, however, can also form from impulsively started flows around two-dimensional bodies, such as cylinders \cite{jeon2004relationship} or airfoils. ~\citet{jeon2004relationship} had employed the impulsive start of a cylinder to study the formation process of the resulting drag vortex pair, which was said to be qualitatively analogous to the vortex ring in ~\citet{gharib1998universal}, except the former was a momentum sink as opposed to the latter, which was a momentum source. The drag vortex pair was found to develop symmetrically at short times for all acceleration values. Past a critical acceleration, it was found that the vortex pair eventually became asymmetric. The starting motion of the airfoil generates an unsteady velocity field, which in turn, causes the stagnation point to form upstream of the trailing edge on the suction side of the airfoil, as opposed to at the trailing edge in a steady field. The stagnation point creates a large velocity gradient around the trailing edge, which further results in the formation of a starting vortex. Owing to Kelvin's circulation theorem, an equal and opposite signed vortex develops around the airfoil, also called the bound vortex. The starting vortex counteracts the circulation over the airfoil, due to its close proximity. As the airfoil surges farther away, its circulation asymptotically tends to the steady state circulation predicted by thin airfoil theory. The starting vortex has been observed and studied extensively following ~\citet{prandtl}'s flow visualisation, both experimentally~\cite{perry, auerbach, whalley_starting} and numerically~\cite{luchini2002start, xu2017computation}. Kaden had modelled the roll-up of a semi-infinite vortex sheet by employing the Birkhoff-Rott equation, and found that the vortex sheet rolls up into a self-similar spiral~\cite{kaden}. The `Kaden problem' was revisited by ~\citet{pullin1}, who employed a numerical approach to solve the Birkhoff-Rott equation, as opposed to Kaden's analytical self-similar roll-up, and reported good agreement between the two methods. Over the years, models predicting vortex sheet roll-up and development have been extended to impulsively started and rotated flat plates with varying velocity profiles~\cite{pullin2021} and arbitrary solid bodies with any number of edges~\cite{pullin_arbit}. Inviscid models, however, do not capture vortex pinch-off and the formation process. Starting vortex roll-up in the case of an impulsively started airfoil is unique from previous studies on vortex formation, since the starting vortex is part of a free vortex sheet, whereas, an equal and opposite circulation resides in the bound vortex, which in turn, is forced to enclose the surging airfoil in order to satisfy the no throughflow condition. 

The objectives of this work are as follows: a) to characterise the unsteady wake generated by an impulsively started airfoil, b) to measure starting vortex evolution whilst it rolls up in the presence of an external stain field of the bound vortex and c) to provide kinematic arguments that explain circulation limits of the starting vortex.  
\begin{figure*}[t!]
    \centering
    \begin{subfigure}{0.45\textwidth}  
        \centering
        \includegraphics[width=\textwidth]{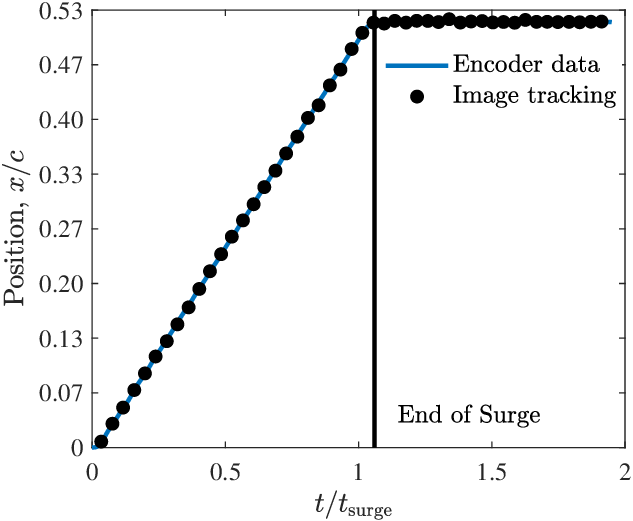}  
        \caption{Airfoil trailing edge position measured by the encoder, as compared to that obtained through image processing.}
        \label{fig:pos}
    \end{subfigure}
    \hfill  
    \begin{subfigure}{0.45\textwidth} 
        \centering
        \includegraphics[width=\textwidth]{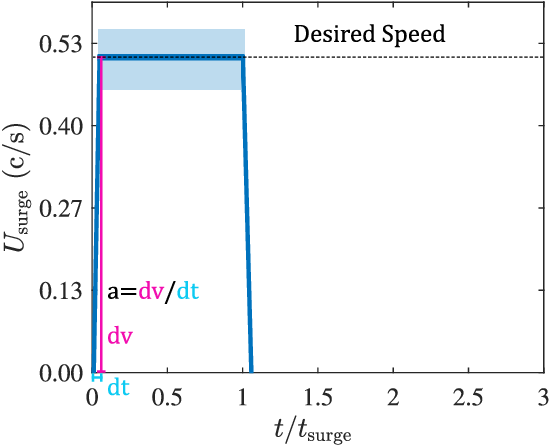}  
        \caption{Velocity profile, computed as a derivative of position. Shaded region demarcates the bounds of velocity fluctuation.}
        \label{fig:vel_profile}
    \end{subfigure}   
    \caption{Position and velocity programme.}
    \label{fig:main_vel}
\end{figure*}
\section{\label{sec:level2}Experimental Setup:\protect\\}
The experiment is designed for a 30 inches long, 12.5 inches wide and 19 inches deep water tank. The vortex-generating airfoil has a NACA0010 profile, with a span of 17.5 inches and a chord size of 2.9 inches, machined in aluminium and anodized. The aspect ratio is 5.8. The blockage ratio, defined as the ratio of the airfoil frontal area to the cross-section area of the test section, is 0.3\%. The wing is attached with end plates on either end to hinder vortex roll-up at the wing tips, thus ensuring that 2D line vortices are generated, similar to the end plates used by ~\citet{leweke1998cooperative}. It is towed through various surge distances along the width of the tank using a Newmark LC Series 300 mm linear traverse, which attaches to the end plates through an attachment plate, 18 inches in span. The spatial accuracy of the traverse is $6\times10^{-4}$ mm/mm of travel. It contains a built in encoder with a sampling frequency of 500 Hz. 
The angle of attack can be altered from $1^\circ$ to $8^\circ$ in $1^\circ$ increments. The velocity field is obtained via time-resolved planar particle image velocimetry, with the field of view illuminated by a light sheet generated using a 1W FN Series Dragon Laser. The light sheet is incident on the at roughly the mid-span of the wing. Images acquired with a Photron Fastcam Mini WX50 camera with a 100 mm lens.  The flow is seeded with 20$\mu$m polyamide particles with a specific gravity of 1.03. Figure~\ref{fig:setup} shows the experimental setup. Images are acquired at 750 fps and post-processed using LaVision's DaVis 10.1 software. Images are preprocessed with a subtracted sliding average filter to minimise noise. A multi-pass variable window size processing method is employed. The first four passes use a 64 $\times$ 64-pixel window with 1:1 square weighting and 50\% overlap. The final pass uses a 24 $\times$ 24-pixel window with 1:1 circular weighting and 50\% overlap. Once the velocity field is obtained, spatial velocity derivatives are obtained using a fourth-order central differencing scheme.

The airfoil is required to start from rest impulsively, hence a velocity program is chosen such that airfoil acceleration is maximised, while maintaining a smooth velocity profile. An example of a motion profile for a constant velocity of 0.5 c/s is shown in Figure~\ref{fig:main_vel}. Figure~\ref{fig:pos} shows the airfoil position measured by the encoder, compared against the trailing edge position obtained through image processing. The two profiles agree well with each other. Figure~\ref{fig:vel_profile} shows the corresponding velocity profile, obtained by taking a central difference of the position obtained by the encoder.

\section{\label{sec:level2}Design Space:\protect\\}
\begin{figure}[t!]
\includegraphics[scale=0.75]{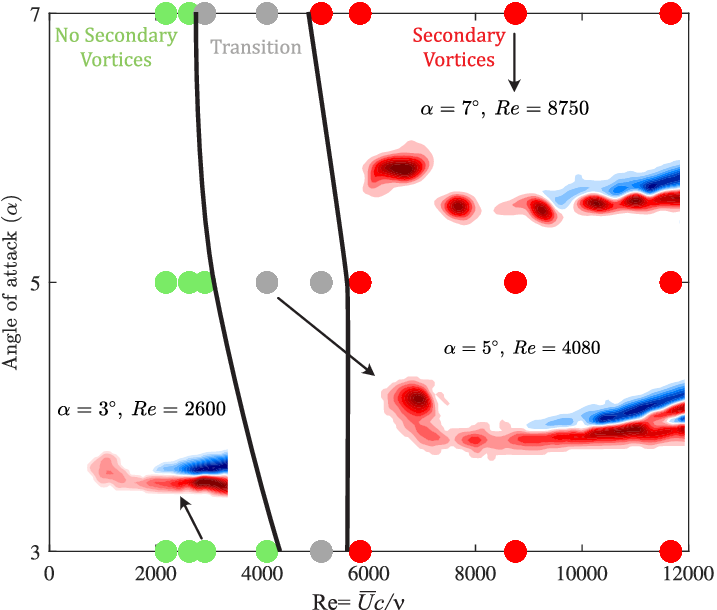}
\caption{Angle of attack versus chord-based Reynolds number, demarcating three flow regimes- attached shear layer, transitional and secondary vortices.}
\label{fig:re_vs_alpha}
\end{figure}

The three initial conditions are airfoil angle of attack ($\alpha$), the distance through which the airfoil surges, ($\Delta x$) and surge speed ($U_\mathrm{surge}=\Delta\dot{x}$). Surge distance was chosen to be half a chord length to avoid the influence of wall effects on starting vortex formation. Wall effects may affect vortex ring formation at a spacing equal to the ring diameter~\cite{walker1987impact}. For the present experiment, assuming that the vortex ring diameter equals the surge distance of 0.5 c, wall spacing is about two times the vortex ring diameter. Consequently, it is assumed that vortex formation and evolution are unaffected by the walls of the water tank.
~\citet{huang_wu_jeng_chen_2001}, who investigated vortex formation and evolution on the suction surface of an impulsively started NACA0010 wing, showed that there are five distinct regimes of vortex evolution based on the angle of attack and chord-based Reynolds number. At angles of attack lower than 10$^\circ$, a separation region was not observed on the suction side for all Reynolds numbers considered. For this study, the highest angle of attack was chosen as $7^\circ$. A lower limit of $3^\circ$ was set, partly due to the weak vorticity that would be produced and the challenges in obtaining accurate measurements. Finally, a range of surge speeds were tested and three distinct regimes were identified for a given angle of attack: one at low surge speeds, where the feeding shear layer remains intact. The second at higher surge speeds, where one part of the shear layer, close to the starting vortex, breaks into secondary vortices while the other remains intact. The third regime is at much higher surge speeds, where the entire shear layer breaks into smaller, Kelvin-Helmholtz-type secondary vortices. Figure~\ref{fig:re_vs_alpha} shows the angle of attack versus the chord-based Reynolds number, where the three regimes are denoted based on secondary vortex formation. Transitional surge speeds can be viewed as critical speeds, $U_{\mathrm{critical}}$, above which the shear layer breaks into secondary vortices. Therefore, a critical Reynolds number can also be defined using this speed and the chord, which decreases as the angle of attack increases. For example, at $\alpha = 3^{\circ}$, $Re_c \approx 4000$, followed by $\alpha=5^{\circ}$, where $Re_{c}\approx 2900$ and at an angle of attack of $7^{\circ}$, the critical Reynolds number was found to be $Re_{c}\approx 2600$.

\section{\label{sec:wake}Wake characteristics:\protect\\ }
\begin{figure}[htbp!]
\includegraphics{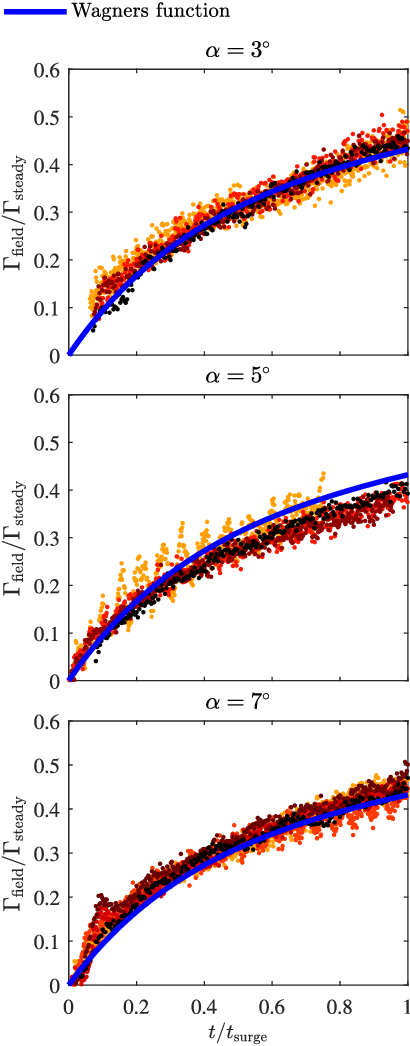}
\caption{Field circulation, $\Gamma_{\mathrm{field}}$, normalised by steady state circulation, $\Gamma_{\mathrm{steady}}$ at angles of attack of $3^{\circ}$, $5^{\circ}$ and $7^{\circ}$ at the following surge speeds- $U_{\mathrm{surge}}= 0.37$ c/s (\cudot{1}),  $U_{\mathrm{surge}}= 0.45$ c/s (\cudot{2}), $U_{\mathrm{surge}}= 0.50$ c/s (\cudot{3}), $U_{\mathrm{surge}}= 0.70$ c/s (\cudot{4}), $U_{\mathrm{surge}}= 1.00$ c/s (\cudot{5}) and $U_{\mathrm{surge}}= 1.50$ c/s (\cudot{6}) . Wagner's function is shown in blue.}
\label{fig:field_wag}
\end{figure}

From Kelvin's circulation theorem, the circulation in the airfoil wake must be equal and opposite to the circulation of the bound vortex enclosing the airfoil. The dynamic response of the fluid to an impulsively started airfoil was first solved analytically by Wagner~\cite{wagner}, who modelled the wake of an impulsive airfoil as a line of point vortices shed from the trailing edge of an impulsively accelerated flat plate, with the last point vortex from the trailing edge being the starting vortex. Wagners' function predicts the lift acting on an impulsively started, thin airfoil at angles of attack lower than steady state stall.~\citet{beckwith2009impulsively} provided an empirical relation for Wagner's function, Equation~\ref{eq:babinsky}, which relates bound circulation to chords travelled by the airfoil, where $\Gamma_{\mathrm{steady}}$ is the circulation predicted by thin airfoil theory and $s^{*}$ is the ratio of surge distance to chord length of the airfoil.    
\begin{equation}
\label{eq:babinsky}
  \Gamma_{\mathrm{b}}/\Gamma_{\mathrm{steady}}= 0.9140-0.3151 e^{-s^{*}/0.1824}-0.5986 e^{-s^{*}/2.0282} 
\end{equation}
Wake circulation, $\Gamma_{\mathrm{field}}$ must, therefore, be equal and opposite to the bound circulation, $\Gamma_{\mathrm{b}}$. Figure~\ref{fig:field_wag} shows the magnitude of wake circulation for three angles of attack, namely, $3^{\circ}$, $5^{\circ}$ and $7^{\circ}$ at various surge speeds, ranging from 0.37 c/s to 1.5 c/s, compared against Wagner's prediction, using Equation \ref{eq:babinsky}, for a surge distance of $0.5$ c. The net circulation in the field agrees well with Wagner's prediction. Consequently, one can speculate that the circulatory lift acting on the airfoil would also be equal to Wagner's lift prediction.

\begin{figure*}[htbp!]
\includegraphics{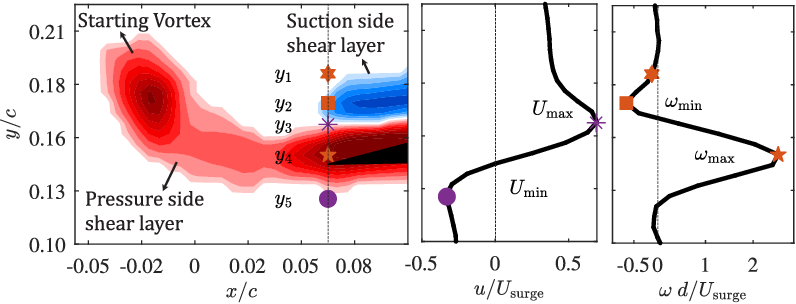}
\caption{Instantaneous vorticity contour at $U_{\mathrm{surge}}= 0.5$ c/s. Also shown are streamwise velocity and vorticity profiles along $y/c$ at a fixed $x/c$ position downstream of the surging airfoil.}
\label{fig:yslice}
\end{figure*}

Apart from the starting vortex, the wake consists of a shear layer shed by the pressure side of the airfoil, with the same sense of vorticity as the starting vortex, and a suction side shear layer with an opposite sense of vorticity, as shown in Figure~\ref{fig:yslice}. At surge speeds where secondary vortices do not form, shear layers can be characterised through velocity and vorticity. Figure~\ref{fig:yslice} shows typical velocity and vorticity profiles along $y/c$ at a fixed downstream location, $x/c= -0.033$ from the airfoil trailing edge. Along $y/c$, the streamwise component of the velocity, $u$, decreases to a local minimum, $U_{\mathrm{min}}$ at the lower edge of the pressure side shear layer, denoted as $y_{5}$. Roughly in the middle of the suction side shear layer, $u/U_{\mathrm{surge}}$ inflects and increases to a local maximum, $U_{\mathrm{max}}$, at $y_{3}$ and finally, tends to zero away from the shear layer. The vorticity reaches a local maximum, $\omega_{\mathrm{max}}$ in the middle of the pressure side shear layer, denoted as $y_{4}$. Further, vorticity inflects and reaches a local minimum, $\omega_{\mathrm{min}}$ at $y_{2}$. At $y_{1}$, which corresponds to the top edge of the suction side shear layer, vorticity begins to tend towards a local minimum. The spatio temporal evolution of $U_{\mathrm{max}}$, $U_{\mathrm{min}}$, $\omega_{\mathrm{max}}$ and $\omega_{\mathrm{min}}$ completely characterises both shear layers. 

\begin{figure*}[t!]
\includegraphics{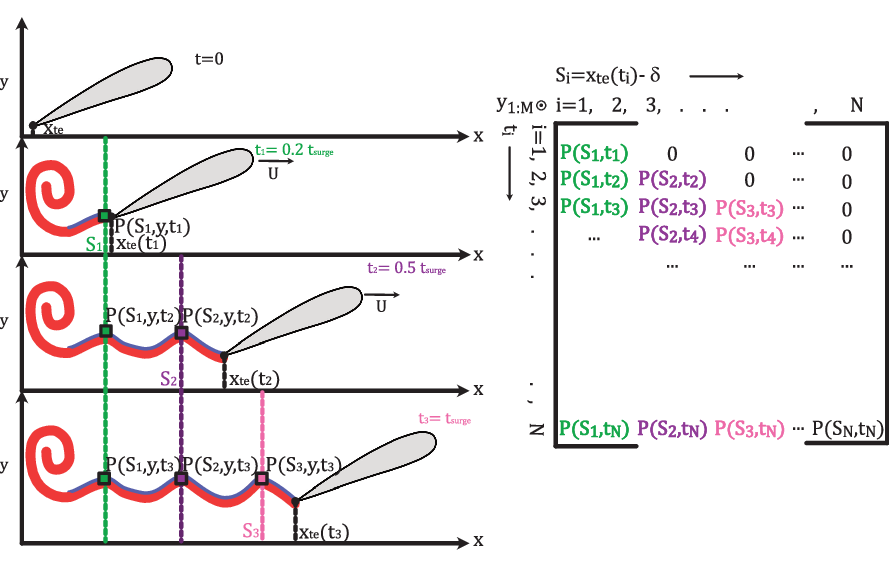}
\caption{Eulerian approach of extracting velocity and vorticity for all $y$ (ranging from 1 to M) at time steps and $x$ spatial locations ranging from 1 to N.}
\label{fig:eulerian}
\end{figure*}
Figure~\ref{fig:eulerian} describes the Eulerian approach that is employed to record velocity and vorticity. Whilst the airfoil is surging, its trailing edge position, $x_{\mathrm{te}}$ is a function of time. At the first time step, $t_{1}$, consider a point, $S_{1}= x_{te}(t_{1})-\textrm{d}x$, where $\textrm{d}x$ is at a small offset behind the airfoil trailing edge (of the order $10^{-1}$ mm). Consider a vector of interest, $P= f(x,y,t)$. At $t= t_{1}$, we extract the value of $P$ at a location $S_{1}=x_{te}(t_{1})-\textrm{d}x\: \forall y$, where $y$ is discretised into $M$ terms. At the next time step, the airfoil trailing edge has passed a point, $S_{2}=x_{te}(t_{2}) -\textrm{d}x$. The quantity, $P$, is now extracted at $x=S_{2}$ as well as $x=S_{1}$ for $t=t_{2}$. As we march forward in time, we establish a location, $S_{i}$ at every time step, which is at a small offset from the trailing edge. At a given time step, $n$, we record the vector $P$ at $S_{n}$, where $i=1, 2, .., n$. Velocity and vorticity along $y$ at every $S_{i}$ position for every time step, $t_{i}$ is recorded using this method. The $x$ position corresponding to the first time step will have the longest time series containing $N$ terms (where $N$ is the total number of time steps during airfoil surge) and the $x$ position corresponding to the last time step will have one data point. This method yields two time series. The first is that of $P$ at the trailing edge whilst the airfoil is surging, \textit{i.e.,} $P(x_{i},y,t_{i})$, where $i=1, 2,..., N$. The second time series is that of how the vector $P$ evolves in time at a given point in space, \textit{i.e.,} $P(x_{i},y,t_{i:N})$. We are only extracting data at $x$ positions excluding the starting vortex and at time steps after starting vortex detachment, which will be presented subsequently. Finally, $S_{1}$ would be the $x$ position closest to the starting vortex and $S_{N}$ would be the $x$ position closest to the airfoil trailing edge at the end of the surge. Once we have recorded the velocity and vorticity values for all $y$ at each $S_{i}$ in time, we can extract the specific values at the five $y$ locations of interest.

\subsection{Streamwise Velocity}
\begin{figure}[htbp!]
\includegraphics[scale=0.9]{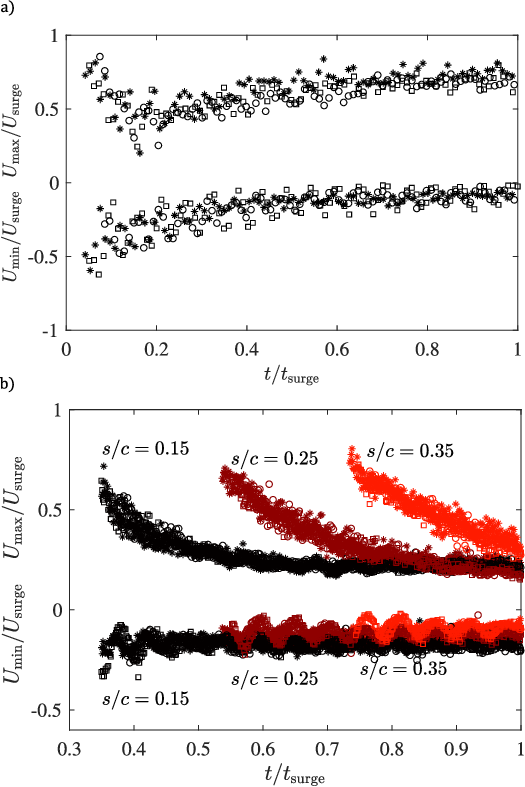}
\caption{a) $U_{\mathrm{max}}/U_{\mathrm{surge}}$ and $U_{\mathrm{min}}/U_{\mathrm{surge}}$ values for three surge speeds a) at the trailing edge whilst the airfoil is surging,  b)  at fixed locations from the starting vortex, namely $s/c=0.15$, $s/c=0.25$ and $s/c= 0.35$. Data are shown for three surge speeds, namely- $U_{\mathrm{surge}}= 0.37$ c/s (*), $0.45$ c/s ($\circ$) and $0.50$ c/s ($\square$).}
\label{fig:umax_all}
\end{figure}

From the streamwise velocity profile shown in Figure~\ref{fig:yslice}, we expect to identify the maximum streamwise velocity, $U_{\mathrm{max}}$ and the minimum streamwise velocity, $U_{\mathrm{min}}$, at each downstream position, $S_{i}=x_{\mathrm{te}}-\textrm{d}x$, at $y_{3}$ and $y_{5}$, respectively. While the airfoil is surging, at a given time step, $t_{i}$, the surge distance is given by $U_{\mathrm{surge}}\times t_{i}$. Thus, we can represent $S_{i}$ by the net airfoil displacement at a given time step, normalised by the airfoil chord length. Figure~\ref{fig:umax_all}a) shows maximum and minimum streamwise velocities normalised by the surge speed, at the trailing edge whilst the airfoil is surging. Since the maximum streamwise velocity acts away from the starting vortex i.e., towards the airfoil, it is representative of the convective speed of the shear layer. In contrast, the minimum streamwise velocity acts towards the starting vortex and is representative of the velocity induced by the starting vortex towards itself. At short surge times, \textit{i.e.,} at $t/t_{\mathrm{surge}}<0.2$, while the airfoil trailing edge is in close proximity of the starting vortex, the magnitude of $U_\mathrm{max}$ and $U_{\mathrm{min}}$ are nearly equal.
As the distance between the starting vortex and airfoil trailing edge increases, the velocity induced by the former decreases proportionally, which results in $U_{\mathrm{min}}$ tending to zero at longer surge times, past $t/t_{\mathrm{surge}}>0.2$. The convective speed of the shear layer, $U_{\mathrm{max}}$ increases with surge time and plateaus to about $0.7\times U_{\mathrm{surge}}$. In a laminar, steady wake, one would expect $U_{\mathrm{min}}$ to be zero, while $U_{\mathrm{max}}$ would maintain a steady value. Through Figure ~\ref{fig:umax_all}a), one can see that as the airfoil surges away from the starting vortex, the wake characteristics tend to that of a steady wake. Figure~\ref{fig:umax_all}b) shows the temporal evolution of maximum and minimum velocities at three fixed points in space: $s/c$= 0.15, 0.25 and 0.35. At a fixed spatial location, the maximum velocity tends to $0.2\times U_{\mathrm{surge}}$ at large surge times, while the minimum velocity tends to about $-0.12\times U_{\mathrm{surge}}$. The surge speed correctly scales both the maximum and minimum velocities in the shear layer.


\subsection{Vorticity}
\begin{figure*}[htbp!]
\includegraphics{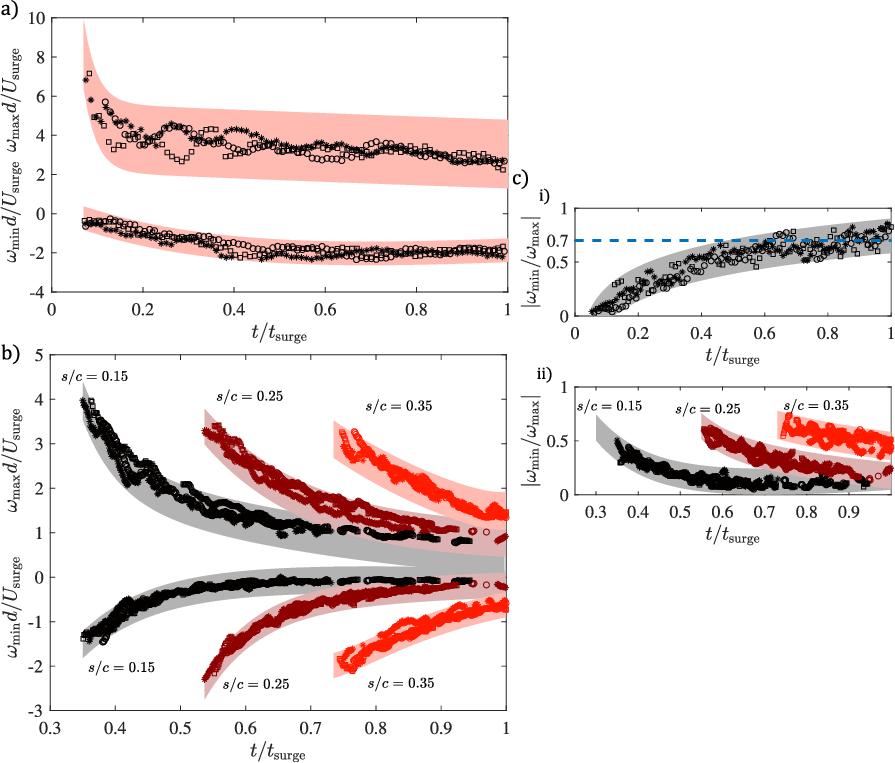}
\caption{a) $\omega_{\mathrm{max}}\: d/U_{\mathrm{surge}}$ and $\omega_{\mathrm{min}}\: d/U_{\mathrm{surge}}$ values for three surge speeds a) at the trailing edge whilst the airfoil is surging  b) at fixed locations from the starting vortex, namely- $s/c=0.15$, $s/c=0.25$ and $s/c= 0.35$. c) Ratio of $\omega_{\mathrm{min}}/\omega_{\mathrm{max}}$ at i)the trailing edge whilst the airfoil is surging and ii) at fixed locations from the starting vortex, namely- $s/c=0.15$, $s/c=0.25$ and $s/c= 0.35$. Shaded regions denote the maximum standard deviation. Data are shown for three surge speeds, namely- $U_{\mathrm{surge}}= 0.37$ c/s (*), $0.45$ c/s ($\circ$) and $0.50$ c/s ($\square$)}
\label{fig:omega_all}
\end{figure*}

Two sets of time series, the maximum and minimum shear layer vorticity, can similarly be extracted, and are shown in Figure~\ref{fig:omega_all}. Vorticity values are scaled with the surge speed and airfoil thickness, $d$. In Figure~\ref{fig:yslice}, it was shown that the minimum vorticity, $\omega_{\mathrm{min}}$ corresponds to the suction side shear layer, while $\omega_{\mathrm{max}}$ corresponds to the pressure side shear layer. From Figure~\ref{fig:omega_all}a, the instantaneous normalised vorticity, $\omega_{\mathrm{max}} d/U_{\mathrm{surge}}$ is maximum at $t/t_{\mathrm{surge}}\approx 0.1$, and tapers to a constant value of 2.8 by the end of surge, \textit{i.e.,} at $t/t_{\mathrm{surge}}= 1$. In contrast, normalised suction side vorticity is relatively smaller at the beginning of surge. At $t/t_{\mathrm{surge}}\approx 0.1$, $\omega_{\mathrm{min}} d/U_{\mathrm{surge}}= -0.3$. Thereafter, the normalised vorticity decreases and plateaus to a steady value of about $-2$ by $t/t_{\mathrm{surge}}=0.4$. 

Normalised vorticity at $s/c=$ 0.15, 0.25 and 0.35 is shown in Figure~\ref{fig:omega_all}b). For the suction side shear layer, it is observed that normalised vorticity decreases over time and tends to a value of 1 at higher times. Minimum vorticity, corresponding to the pressure side shear layer, initially has a value of $-2$ and tends towards $-0.1$. Figure~\ref{fig:omega_all}c) shows the relative strength of the two shear layers at the trailing edge as well as at fixed spatial locations. The ratio of $\omega_{\mathrm{min}}/\omega_{\mathrm{max}}$ is always less than 1, which shows that the pressure side shear layer is stronger than the suction side. At the trailing edge, the ratio increases from nearly 0 to 0.7 at $t/t_{\mathrm{surge}}= 0.4$ and plateaus thereafter. At fixed spatial locations, $\omega_{\mathrm{min}}/\omega_{\mathrm{max}}$ tends to 0 at larger times, which is a consequence of $\omega_{\mathrm{min}}$ tending to zero. 

The wake of an impulsively started airfoil has been characterised by net circulation and spatio-temporal evolution of velocity and vorticity. Starting vortex characteristics are presented in the following section.

\section{Starting Vortex Circulation}
Using the $\lambda_{\mathrm{ci}}$ criterion, the starting vortex is identified and isolated from both shear layers. Also known as the swirl strength, the $\lambda_{\mathrm{ci}}$ criterion has been used in several studies to distinguish between regions of high shear and coherent vortices, where vorticity contours fail to demarcate the extent of a vortex~\cite{baskaran2022lagrangian, francescangeli2021discrete}. 

\begin{figure*}[t!]
\includegraphics{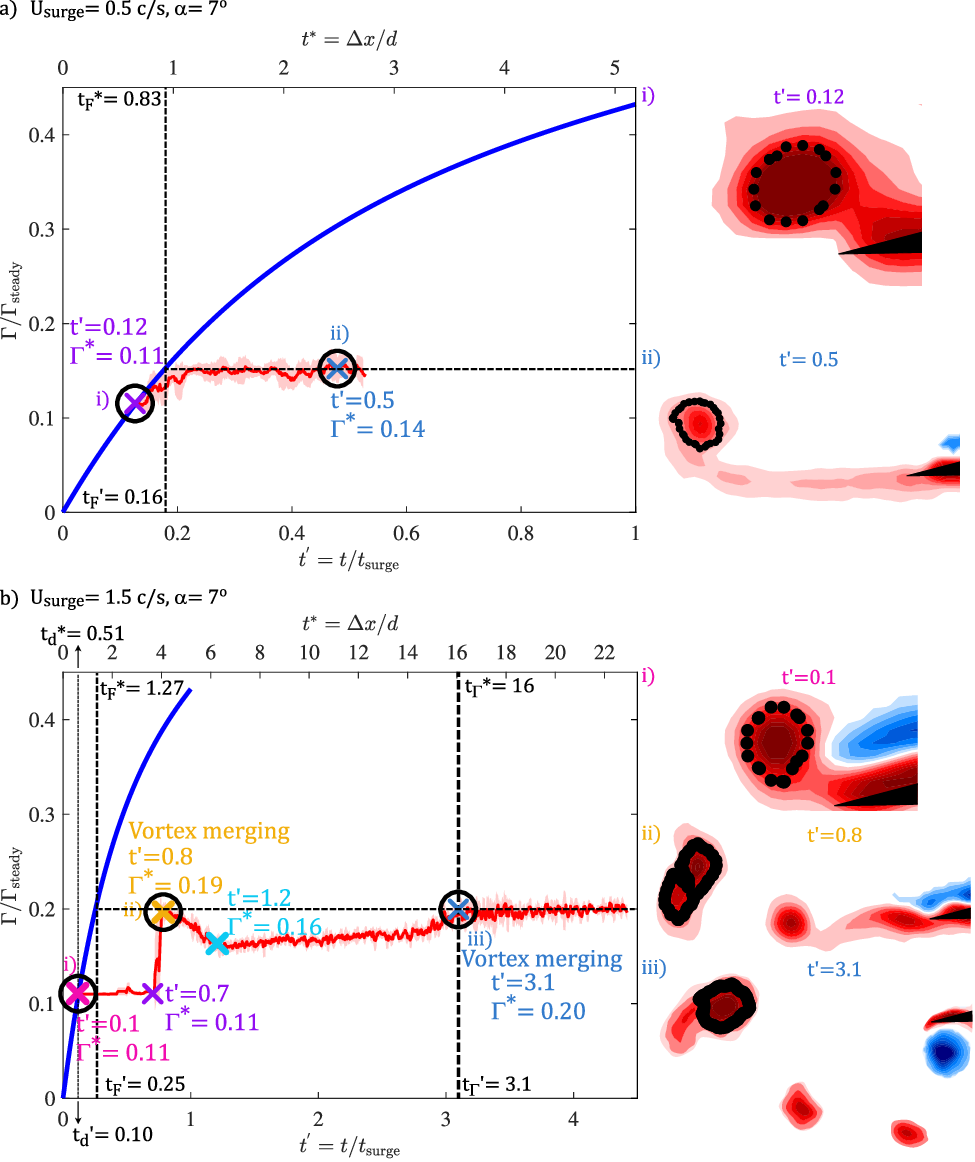}
\caption{ Starting vortex circulation, $\Gamma$, normalised by circulation obtained from thin airfoil theory for two cases- a) $U_{\mathrm{surge}}< U_{\mathrm{critical}}$ and b) $U_{\mathrm{surge}}> U_{\mathrm{critical}}$. The extent of the starting vortex is defined by using the $\lambda_{\mathrm{ci}}$ criterion. Wagner's function ({\protect\coline{7}}) is superimposed.}
\label{fig:circ2}
\end{figure*}

In vortex formation studies, it is common practice to compare the net circulation in the field with that of the primary vortex~\cite{gharib1998universal, o2014pinch, limbourg2021formation, karen_conical}. The formation number is defined as the slug flow time corresponding to the circulation generated by the vortex generator, beyond which excess circulation is rejected by the primary vortex ring. Pinch-off is identified as the instant at which primary vortex circulation ceases to increase despite the presence of excess vorticity generated by the apparatus, in this case, the surging airfoil. In ~\citet{gharib1998universal}'s work, as well as studies that followed, it was argued that primary vortex circulation ceases to increase once the pinch-off process is complete, resulting in the accumulation of excess vorticity within the trailing jet, which in turn, rolls up to form secondary, Kelvin-Helmholtz type vortices~\cite{jeon2004relationship,ringuette2007role,o2014pinch}. The time at which the vortex achieved its maximum circulation coincided with the vortex detaching from its feeding shear layer, shortly followed by secondary vortex formation in the shear layer. The simultaneity of all three events masked cause and effect. Although ~\citet{gharib1998universal} had cautioned that pinch-off is not a discrete process and can take up to two non-dimensional time units for completion, the definition of pinch-off time encompassed primary vortex detachment from its trailing jet and the formation of secondary vortices. ~\citet{limbourg2021formation} studied vortex ring from an orifice, and observed that the primary vortex ring detaches from the vortex generating apparatus at a `detachment time'. Subsequently, secondary vortices were released by the vortex generator. At a limiting stroke ratio (or $L/D$), the detached primary vortex subsequently merges with trailing secondary vortices and attains a saturation in circulation, indicating the completion of the formation process in the sense proposed by ~\citet{gharib1998universal}, where primary vortex ring circulation ceases to grow. ~\citet{limbourg2021formation} had further defined a maximum circulation formation time, $t^{*}_{\Gamma}$ which marks the end of primary vortex merger with trailing secondary vortices. Their observations highlighted that the primary vortex can detach from its feeding shear layer without attaining saturation in circulation. Furthermore, secondary vortex formation is not caused strictly due to pinch-off, rather it can occur due to shear layer instabilities. 

Two non-dimensional times are considered. The first is an analogous slug flow time, $t^{*}= \overline{U}_{p}\: t/d$, which would simplify to $\Delta x_{\mathrm{surge}}/d$, where $d$ is the airfoil thickness. This definition is consistent with previous experiments, such as that of an impulsively started cylinder, where non-dimensional time was defined as the ratio of surge distance to cylinder diameter~\cite{jeon2004relationship}. The second non-dimensional time, $t'$ is obtained by normalising time with surge time, $t_{\mathrm{surge}}$, consistent with the work of ~\citet{devorianringuette} on an impulsively rotated, low aspect ratio trapezoidal plate. Figure~\ref{fig:circ2} shows the temporal evolution of a starting vortex for two cases, one at $U_{\mathrm{surge}}= 0.5$ c/s, where secondary vortices do not form and another at $U_{\mathrm{surge}}= 1.5$ c/s, where the shear layer breaks into secondary vortices. 
In the former, starting vortex circulation is equal to the net circulation in the field at short non-dimensional surge times, \textit{i.e.,} at $t'<0.16$. Vorticity contours at $t'= 0.12$ show a nascent starting vortex rolling up at the airfoil trailing edge. By $t'= 0.16$, non-dimensional circulation, $\Gamma^{*} = \Gamma/\Gamma_{\textrm{steady}}$, attains a value of 0.14, which remains constant throughout the time series shown here. At half the surge time, $t'= 0.5$, $\Gamma^{*}$ still equals 0.14, despite the presence of excess vorticity contained within the shear layer, visualised through vorticity contours. In this case, the `detachment time', $t'_{d}$ defined by ~\citet{limbourg2021formation}, where the starting vortex is no longer attached to the feeding shear layer, corresponds to pinch-off time, where the starting vortex rejects additional circulation in the field, which equals to $t'=0.16$. The formation time, \textit{i.e.,} the time corresponding to the maximum circulation attained by the starting vortex, $t'_{F}$ is also 0.16. Furthermore, ~\citet{gharib1998universal} had observed that starting vortex detachment from its feeding shear layer results in secondary, Kelvin-Helmholtz-type vortices. From the vorticity contours shown in Figure~\ref{fig:circ2}a)ii), one can see that the shear layer remains intact despite starting vortex detachment, indicating that secondary vortex formation is not solely a consequence of primary vortex detachment. In terms of slug flow time, $t^*$, the formation time is $t_{F}^{*}=0.83$, while the detachment time is $t_{d}^{*}\approx 1$.

In the second case, where $U_{\mathrm{surge}}= 1.5$ c/s, $\Gamma^{*}$ remains nearly constant at a value of 0.11 from $0.1<t/t_{\mathrm{surge}}<0.7$. The slug flow detachment time in this case would be ${t^{*}}_{d}= 0.51$, while $t'_{d}= 0.1$. The vorticity contours at $t'= 0.1$ show the starting vortex rolling up at the airfoil trailing edge. Between $0.7<t/t_{\mathrm{surge}}<0.8$, the starting vortex merges with a neighbouring secondary vortex, visualised through the vorticity contours shown at $t'= 0.8$, where the $\lambda_{\mathrm{ci}}$ criterion identifies the primary vortex as a contour enclosing both the starting and first secondary vortex. Consequently, $\Gamma/\Gamma_{\mathrm{steady}}$ increases from 0.11 to 0.19. At $0.8<t/t_{\mathrm{surge}}<1.2$ non-dimensional starting vortex circulation decreases from 0.19 to 0.16, likely due to viscous dissipation as a result of the merging process. Thereafter, starting vortex circulation remains nearly constant up until $t'= 3.1$. At $t/t_{\mathrm{surge}}= 3.1$, the primary vortex once again begins to merge with a neighbouring secondary vortex, resulting in  $\Gamma/\Gamma_{\mathrm{steady}}$ increasing from 0.16 to 0.20. The vorticity contour at this time instant shows a secondary vortex near the starting vortex. The $\lambda_{\mathrm{ci}}$ criterion mostly encloses the primary vortex, unlike the vorticity contour shown at $t'= 0.8$, which encompasses both the primary and first secondary vortex. For this case, $t^{*}_{F}$ as defined by ~\citet{gharib1998universal} would be 1.27 and ${t_{\Gamma}}^{*}$, defined by ~\citet{limbourg2021formation} would be 16. The corresponding non-dimensional times using $t_{\mathrm{surge}}$ would be $t'_{F}= 0.25$ and $t'_{\Gamma}= 3.1$.

\begin{figure}[t]
\includegraphics{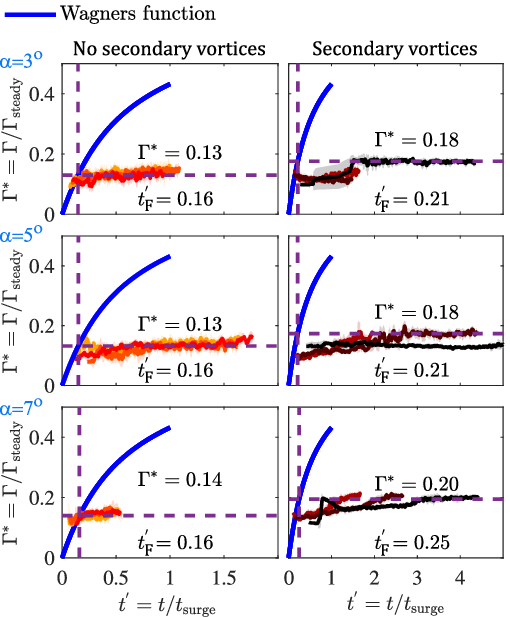}
\caption{Starting vortex circulation, $\Gamma$, normalised by steady state circulation, $\Gamma_{\mathrm{steady}}$ at angles of attack of $3^{\circ}$, $5^{\circ}$ and $7^{\circ}$ at the following surge speeds- $U_{\mathrm{surge}}= 0.37$ c/s ({\protect\coline{1}}),  $U_{\mathrm{surge}}= 0.45$ c/s ({\protect\coline{2}}), $U_{\mathrm{surge}}= 0.50$ c/s ({\protect\coline{3}}), $U_{\mathrm{surge}}= 0.70$ c/s ({\protect\coline{4}}), $U_{\mathrm{surge}}= 1.00$ c/s ({\protect\coline{5}}) and $U_{\mathrm{surge}}= 1.50$ c/s ({\protect\coline{6}}). Also plotted is Wagner's function ({\protect\coline{7}}).}
\label{fig:tprime}
\end{figure}

The results suggest two distinct formation processes, one for cases when there are no secondary vortices, and another when there are i.e., $Re>Re_c$. The analysis is extended to other angles of attack and shown in Figure~\ref{fig:tprime}for a surge distance of 0.5 c. Since the non-dimensional circulation in the field agrees with Wagner's prediction, the value of $\Gamma^{*}$ at the end of surge for every case presented in Figure~\ref{fig:tprime} is equal to 0.43. 
For each angle of attack, at surge speeds where secondary vortices do not form, \textit{i.e.,} $U_{\mathrm{surge}}< 0.50$ c/s, non-dimensional circulation plateaus to a value of about 0.13, which is $\approx 33\%$ of net field circulation, at a non-dimensional time of 0.16. At higher surge speeds, $\Gamma^{*}$ attains a value of $0.18-0.2$, which is approximately $42-47\%$ of the total circulation in the field. Due to the consistency in the value of $\Gamma^{*}$, one may speculate that starting vortex circulation can be explained by the formation number argument, where the starting vortex is unable to absorb additional circulation. However, with the higher surge speeds for the same angle of attack, the starting vortex subsequently absorbs the circulation from the secondary vortex, implying it could not have attained its maximum value before separating from the shear layer. In other words, the starting vortex is not fully formed when it separates from the shear layer; hence the appearance of the secondary vortices in the shear layer is not a formation-type related problem. Starting vortex detachment from the feeding shear layer, as well as merging neighbouring secondary vortices corroborates the work of ~\citet{limbourg2021formation}. These observations emphasize that starting vortex detachment and secondary vortex formation do not necessarily occur simultaneously and aren't strictly an effect of pinch-off. Another noteworthy observation is that a primary vortex can detach from its feeding shear layer despite not being fully formed. 

From Figure~\ref{fig:circ2}, each kinematic event, namely starting vortex detachment and merging with neighbouring secondary vortices, can be associated with a slug flow time, denoted using $t^{*}$ as well as a non-dimensional time normalised by $t_{\mathrm{surge}}$, denoted by $t'$. The final circulation of the starting vortex can also be associated with two non-dimensional formation time values. Table~\ref{tab:t_table} summarises the results for the circulation time series shown in Figure~\ref{fig:tprime}.
\begin{table*}
\caption{\label{tab:t_table} Summary of normalised time values for cases where secondary vortices do not form ($Re<Re_{c}$) and where the shear layer breaks into secondary vortices ($Re>Re_{c}$)}
\begin{ruledtabular}
\begin{tabular}{cccc}
    Normalising Parameter & Quantity & {${Re<Re_c}$} & {${Re>Re_c}$} \\ 
    \hline
    \hline
    \multirow{3}{*}{$t_{\mathrm{surge}}$} 
    & Detachment Time, $t'_{d}$ & 0.16 & 0.10 \\
    & Maximum circulation formation time, $t'_{\Gamma}$ & 0.16 & 1.5-3.1 \\
    & Formation time, $t'_{F}$ & 0.16 & 0.21-0.25 \\
\hline
        \multirow{3}{*}{$d/\overline{U}_{p}$} 
    & Detachment Time, $t^{*}_{d}$ & 0.78-0.95 & 0.5-0.75 \\
    & Maximum circulation formation time, $t^{*}_{\Gamma}$ & 0.78-0.95 & 7.8-15 \\
    & Formation time, $t^{*}_{F}$ & 0.78-0.95 & 1.1-1.3 \\
    
\end{tabular}
\end{ruledtabular}
\end{table*}

Since the limit on starting vortex circulation is not due to the formation number argument, it is more likely that the starting vortex forms due to it detaching from the shear layer, similar to what is observed in vortex ring formation from orifices\cite{limbourg2021formation} and rotating plates \cite{francescangeli2021discrete}. We postulate that detachment occurs due to a competition between the velocity induced by the nascent, clockwise starting vortex towards itself and the tendency of the shear layer to convect towards the bound vortex, enclosing the airfoil, which is away from the starting vortex. In the following section, this is demonstrated both, qualitatively and quantitatively. 

\section{Finite Time Lyapunov Exponents}
FTLEs indicate regions of maximum particle separation over a finite integration time, $T$. Particle separation is measured through the Jacobian of the flow map, $\phi^{t_{o}+T}_{t_{o}}$. Maximum eigenvalues of the Jacobian, $\lambda_{\mathrm{max}}$, are used to compute the Finite Time Lyapunov Exponent, a scalar quantity, $\sigma$, given by Equation~\ref{eq:ftle2}.
\begin{equation}
\sigma(\phi^{t_{o}+T}_{t_{o}})=\frac{1}{|T|}\: \mathrm{log} \sqrt{\lambda_{\mathrm{max}}}
\label{eq:ftle2}
\end{equation}
Regions of local maxima, called ridges, are analogous to separatrices in non-linear dynamic systems, and demarcate dynamically distinct regions in the flow. The integration time, $T$ can be either positive (pFTLEs), in which case ridges denote maximum particle separation, or negative (nFTLEs), where ridges denote maximum particle attraction. 

\begin{figure*}[t!]
\includegraphics{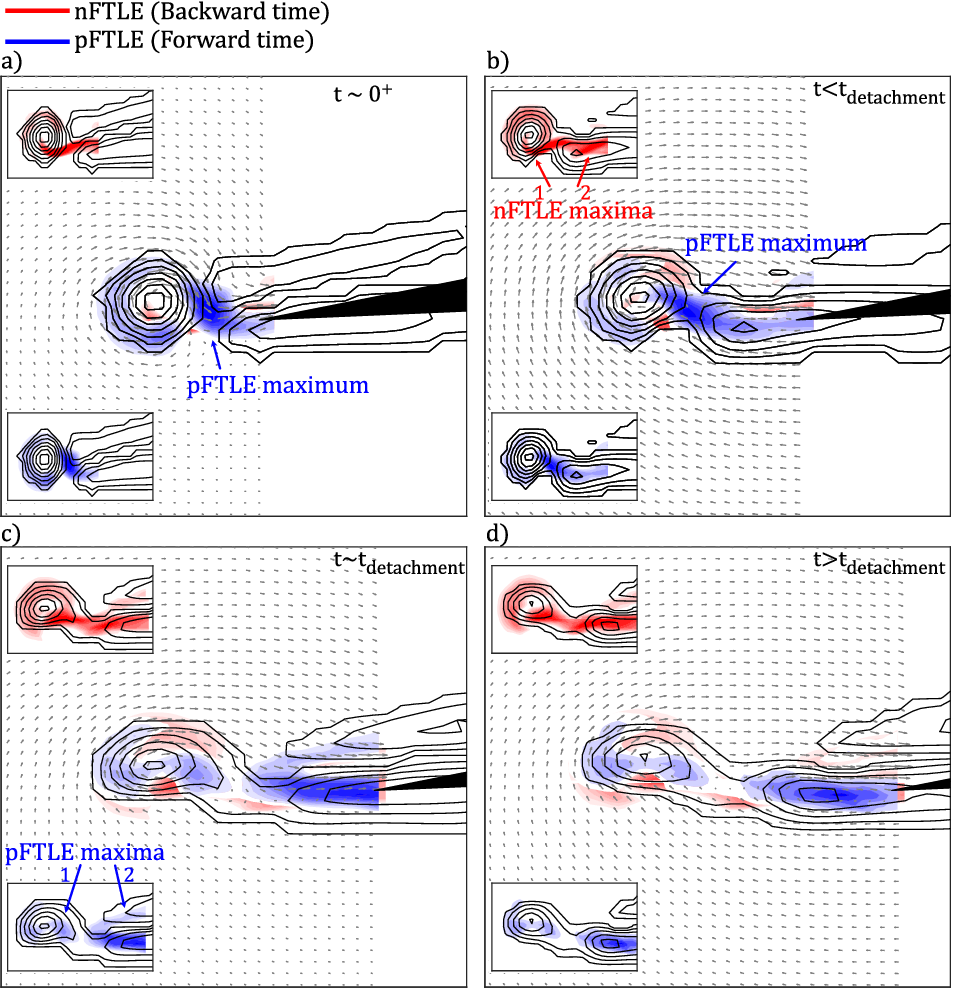}
\caption{Forward and negative time FTLEs superimposed with vorticity contours and velocity vector at four times- a) shortly after surge, b) before detachment, c) at detachment and d) several time steps after detachment. Inset are individual negative and positive FTLE contours.}
\label{fig:ftle}
\end{figure*}

FTLEs are computed by averaging particle trajectories, thereby serving as a reliable visual tool at time steps leading to detachment, where the starting vortex is attracting particles towards itself, whilst the bound vortex enclosing the airfoil is pulling fluid particles in the opposite direction, away from the starting vortex. One would thus expect to see positive FTLE ridges between the starting vortex and the shear layer close to detachment, while negative FTLE ridges would be visible within the starting vortex and shear layer. Furthermore, FTLEs are preferred to visualise detachment over vorticity contours, since the latter can be misleading. For example, in Figure~\ref{fig:circ2}, the vorticity contour shown for $U_{\mathrm{surge}}= 1.5$ c/s, at $t'= 3.1$ appears as though the primary vortex has merged with a neighbouring secondary vortex, while the $\lambda_{ci}$ criterion does not encompass both vortices. The circulation value at this time step does not increase significantly, as compared to the increase at $t'= 0.7$, where vortex merging first takes place. Another way of visualising the field is through streamlines, which was used by ~\citet{gan2012drag} to identify the extent of vortical structures and used by ~\citet{sattari2012growth} to demonstrate detachment. Interpreting streamlines for unsteady flows can be erroneous, especially for identifying critical points, since streamlines are computed using the instantaneous velocity field and there is no correlation between consecutive snapshots in time. Consequently, what appears as a critical point at one time step may not be a critical point at the next time step. FTLEs, on the other hand, are identified based on particle trajectories over a finite time, thereby mitigating the errors introduced by using instantaneous vorticity and streamlines.

FTLEs have been successfully used to identify structures in vortex-dominated flows~\cite{shadden2006lagrangian, shadden2007transport, green2007detection,o2010lagrangian, baskaran2022lagrangian}. There are two considerations: the integration time, $T$ and the spatial resolution. To avoid long computation time, the integration time is chosen as the shortest time for which resulting FTLEs capture the boundaries of the vortex. ~\citet{o2014pinch} had used an integration length slightly shorter than the formation time for visualising detachment. For the FTLE plots presented here, the integration length equals $0.5 \times t_{\mathrm{detachment}}$ of the starting vortex. Similar considerations apply to the choice of grid size. ~\citet{o2014pinch} had employed a resolution of $0.01\times D$, where $D$ is the natural length scale of the vortex generator. In the present study, the natural length scale would be the thickness of the airfoil, $d= 7.5$ mm. Thus, the resolution in $x$ and $y$ was fixed as 0.07 mm. 

Figure~\ref{fig:ftle} shows positive and negative time FTLEs superimposed with vorticity contours and velocity vectors, at four time intervals. Shortly after the airfoil begins to surge, at \textit{i.e.,} $t\sim0^{+}$, the pFTLE encloses the starting vortex. The region between the starting vortex and the airfoil trailing edge shows a higher value of the pFTLE, as compared to the rest of the starting vortex. This highlights the effect of vortex roll-up in the absence of rotational symmetry, since one would expect pFTLE ridges enclosing a symmetric primary vortex to be nearly equal in magnitude. The higher pFTLE value at the junction between the starting vortex and the shear layer demonstrates that particle trajectory at short time scales, of the order of vortex detachment, is dictated by the asymmetry. The nFTLE shows a local maxima between the starting vortex centre and the nascent shear layer, indicating regions of high particle attraction. During roll-up, at $t<t_{\mathrm{detachment}}$, the pFTLE ridge continues to enclose the starting vortex, albeit with a higher value at the junction between the starting vortex and shear layer. Backward-time, or nFTLE ridges, also continue to enclose the starting vortex, in addition to showing a higher value at the starting vortex and shear layer junction (denoted as maxima 1), as well as the middle of the shear layer, closer to the trailing edge (denoted maxima 2). Since an FTLE is a scalar quantity, the maxima indicate that both the starting vortex and shear layer are regions of high particle attraction. However, the maxima do not indicate the direction of particle attraction and therefore, do not show whether the two structures are competing against each other. The pFTLE maxima, which lie between the two structures, indicate that the junction is a region of high particle separation. Thus, one can speculate that nFTLE maxima 1, which lies on one end of the pFTLE maximum, entrains fluid particles in one direction, that of the starting vortex, whereas nFTLE maxima 2, which lies within the shear layer, attracts particles towards itself in the opposite direction, competing against the starting vortex. The points of intersection of forward and backward time FTLEs indicate saddles, which have been used in previous studies to identify vortex detachment~\cite{mulleners2012onset, krishna2018flowfield}. As mentioned previously, vortex detachment is not a discrete process and thus we interpret the emergence of a saddle point as an indicator of the onset of vortex detachment. At detachment, \textit{i.e.,} $t\sim t_{\mathrm{detachment}}$, we observe two distinct, disconnected pFTLE ridges, one associated with the starting vortex, indicated as pFTLE maxima 1, and another with the shear layer, labelled as pFTLE maxima 2 in Figure~\ref{fig:ftle}. At later time steps, $t>t_{\mathrm{detachment}}$, the two forward-time FTLE ridges continue to be disconnected. The emergence and continued appearance of disconnected pFTLE ridges indicate that both flow features associated with a pFTLE ridge are dynamically distinct, thereby, indicating that the starting vortex has detached from its feeding shear layer. 

It is also observed that pFTLE ridges in ~\ref{fig:ftle}c) and d) do not fully enclose the starting vortex. One could speculate that this occurs because of the choice of integration time, $T$. Longer integration times yield fine, distinct FTLE ridges. However, increasing the integration time by 3 times its present value of $0.5\times t_{\mathrm{detachment}}$ yielded similar pFTLE ridges that did not fully enclose the starting vortex at time steps leading to and after detachment. It is conjectured that the asymmetry in starting vortex roll-up results in uneven velocity gradients, with higher velocity gradients existing at the airfoil trailing edge. Consequently, pFTLE ridges, which indicate regions of maximum particle separation may favour some parts of the vortex over others, which results in partial enclosure of the starting vortex by the corresponding pFTLE ridge. In contrast, the nFTLE ridge, which occurs in regions that attract fluid particles in backward-time, fully encloses the starting vortex up until its detachment from the shear layer. Since the starting vortex attracts fluid particles regardless of symmetry, it is unsurprising that nFTLEs fully enclose the starting vortex.  

Through FTLEs, regions and time scales where starting vortex roll-up competes with shear layer convection have been identified. In the following section, the competition between the two structures is quantified.

\section{Kinematics leading to detachment}
\begin{figure*}[htbp!]
\includegraphics{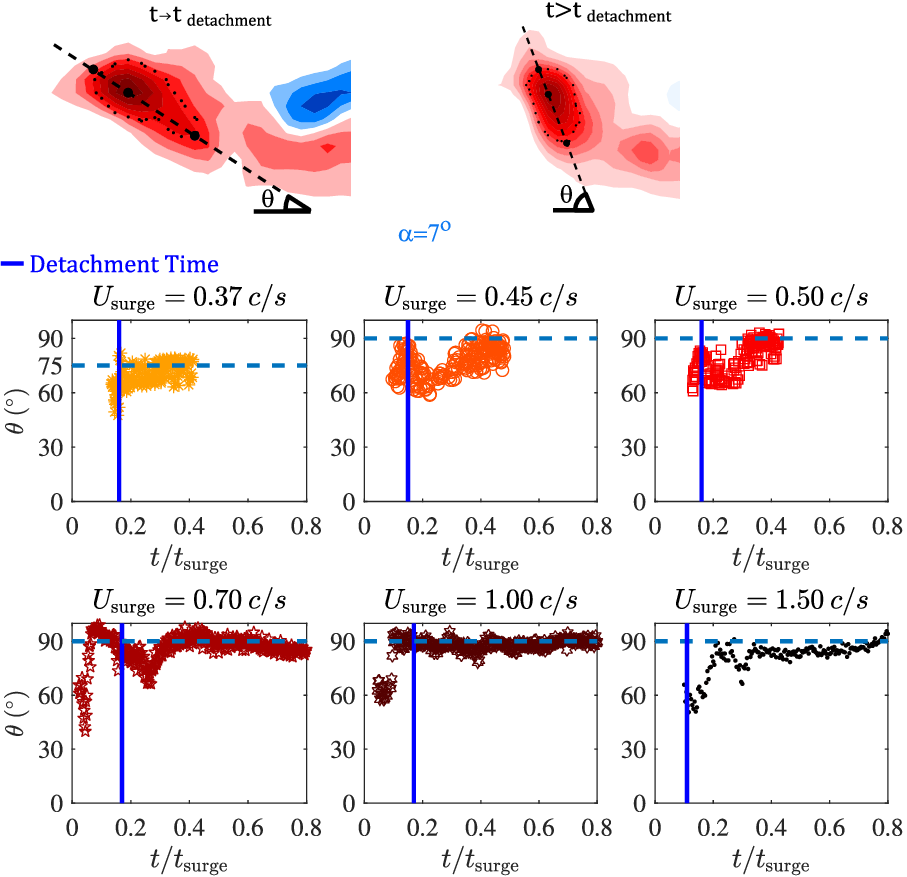}
\caption{Starting vortex tilt angle, $\theta$ at an angle of attack of $7^{\circ}$ for $U_{\mathrm{surge}}= 0.37$ c/s (\cudot{1}),  $U_{\mathrm{surge}}= 0.45$ c/s (\cudot{2}), $U_{\mathrm{surge}}= 0.50$ c/s (\cudot{3}), $U_{\mathrm{surge}}= 0.70$ c/s (\cudot{4}), $U_{\mathrm{surge}}= 1.00$ c/s (\cudot{5}) and $U_{\mathrm{surge}}= 1.50$ c/s (\cudot{6}).  }
\label{fig:theta}
\end{figure*}

Shortly after the airfoil begins to surge, the starting vortex, elliptical in shape, rolls up at the airfoil trailing edge. The major principal axis, defined as the longest diameter of the vortex, makes an angle, $\theta$ with the $x$ axis in a Cartesian coordinate system, shown in Figure~\ref{fig:theta}. In general, $\theta$ increases from roughly $50^{\circ}-60^{\circ}$ to $90^{\circ}$ at short times. Thereafter, the tilt angle decreases to about $60^{\circ}$ and eventually asymptotes to $90^{\circ}$ for all surge speeds, with the exception of $U_{\mathrm{surge}}= 0.37$ c/s, where the value tends to $75^{\circ}$. One can speculate that the initial increase in $\theta$ to $90^{\circ}$ is due to the proximity of the airfoil to the starting vortex, which imposes an asymmetric strain field on the latter and governs its short-time dynamics and structure. The initial elliptic shape of the starting vortex is reminiscent of vortical streamlines in external strain fields. With an increase in distance between the surging airfoil and the starting vortex, the effect of external strain on the starting vortex can be expected to diminish.  

\begin{figure*}[ht]
\includegraphics{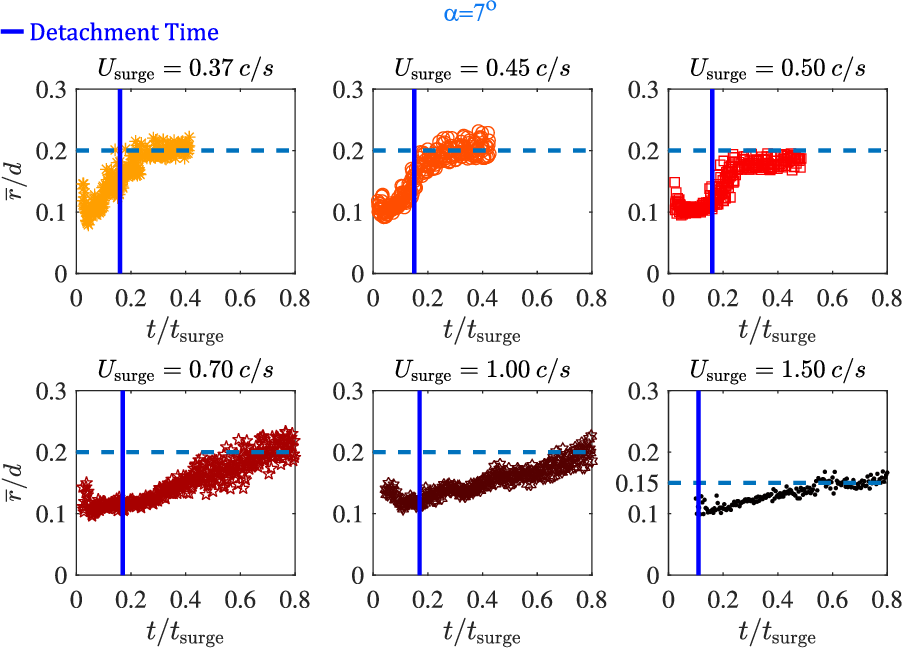}
\caption{Mean starting vortex radius, $\overline{r}$, normalised by airfoil thickness, $d$ at an angle of attack of $7^{\circ}$ for $U_{\mathrm{surge}}= 0.37$ c/s (\cudot{1}),  $U_{\mathrm{surge}}= 0.45$ c/s (\cudot{2}), $U_{\mathrm{surge}}= 0.50$ c/s (\cudot{3}), $U_{\mathrm{surge}}= 0.70$ c/s (\cudot{4}), $U_{\mathrm{surge}}= 1.00$ c/s (\cudot{5}) and $U_{\mathrm{surge}}= 1.50$ c/s (\cudot{6})}
\label{fig:rmean}
\end{figure*}

An average vortex radius, $\overline{r}$, is defined as the mean distance between the vortex centre and each point of the contour enclosing the starting vortex. Figure~\ref{fig:rmean} shows $\overline{r}$ normalised by airfoil thickness, $d$. At surge speeds lower than 0.7 c/s, the average vortex size increases whilst the airfoil surges and plateaus to approximately $0.2\:d$ shortly after detachment. As the surge speed increases to cases where secondary vortices form, \textit{i.e.,} at $U_{\mathrm{surge}}> 0.5$ c/s, the average vortex radius continues to increase after starting vortex detachment due to merging with neighbouring secondary vortices. The plateau in $\overline{r}/d$ is observed at higher $t/t_{\mathrm{surge}}$ of 0.6-0.8, as opposed to 0.2-0.3 for lower surge speeds, without secondary vortex formation. With the exception of $U_{\mathrm{surge}}= 1.50$ c/s, where $\overline{r}/d$ asymptotes to 0.15, the normalised mean vortex radius tends to 0.2.     

\begin{figure*}[htbp!]
\includegraphics{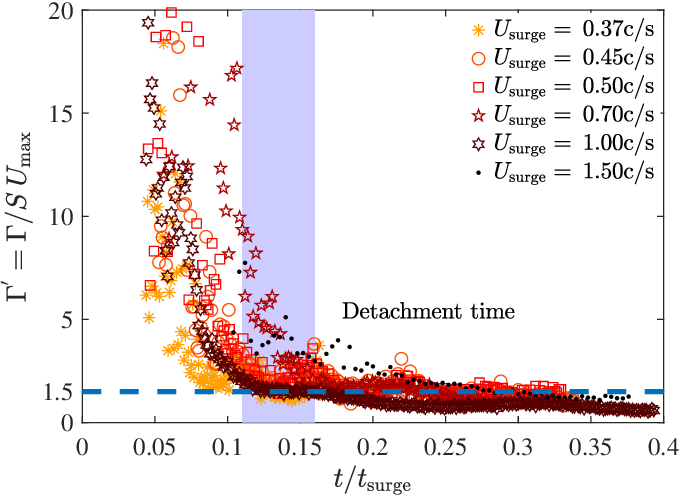}
\caption{Detachment parameter, $\Gamma'$, proposed by ~\citet{sattari2012growth} at an angle of attack of $7^{\circ}$ for various surge speeds.} 
\label{fig:gamma_star}
\end{figure*}

~\citet{sattari2012growth} had studied starting vortex formation from an impulsively started jet intending to describe vortex pinch-off in the absence of a natural length scale of a vortex generator, such as the diameter of the cylinder. They had explained that pinch-off occurs due to a competition between the inertia of the shear layer that forces it to remain in the streamwise direction and the velocity induced by the starting vortex. They had defined a non-dimensional parameter, $\Gamma'=\Gamma/SU_{\mathrm{max}}$, where $\Gamma/S$ is the starting vortex induced velocity at a distance $S$ from its centre and $U_{\mathrm{max}}$ is the maximum convective speed of the shear layer. At pinch-off, $\Gamma'$ tended to a value of 1.5. For this study, the same parameter is computed using the measured starting vortex circulation and maximum streamwise velocity, $U_{\mathrm{max}}$. Figure~\ref{fig:gamma_star} shows the parameter, $\Gamma'$ at various surge speeds for an angle of attack of $7^{\circ}$. The detachment time, $t'_{d}$ lies between 0.1 to 0.16 for all Reynolds numbers considered here, as shown in Table~\ref{tab:t_table}. Consistent with ~\citet{sattari2012growth}, the non-dimensional parameter converges to a value of 1.5. Similar results were obtained for angles of attack of $3^{\circ}$ and $5^{\circ}$, although not shown here for brevity. This, in turn, proves that starting vortex detachment, for the set of initial conditions considered here, can be explained through a kinematic argument, instead of the formation number argument. As such, if a limiting process exists for this specific problem, it occurs at higher surge speeds or angles of attack. This also proves that a limiting process would result in detachment, however, detachment does not necessarily imply the occurrence of a limiting process that results in a saturation of vortex circulation. Finally, the results also suggest that vortex detachment is not necessarily followed by secondary vortex formation. Thus, all kinematic events that were observed to occur simultaneously in some of the previous studies on vortex formation~\cite{gharib1998universal, jeon2004relationship, ringuette2007role}, occur discretely in the case of an impulsively started airfoil, which lacks rotational symmetry. The starting vortex, which has not attained its maximum circulation, detaches from the feeding shear layer, which, in turn, may or may not form secondary vortices, depending on surge speed.

\section{Conclusions}
The starting vortex and shear layers generated by the impulsive start of a NACA0010 airfoil were measured and analysed. The design space consisted of angles of attack ranging from $3^{\circ}$ to $7^{\circ}$ and surge speeds ranging from $0.37$ c/s to $1.5$ c/s. For a given angle of attack, the shear layer remained intact up to a critical surge speed, beyond which the shear layer broke into a train of Kelvin-Helmholtz-type vortices. The critical surge speed was found to be a function of the angle of attack, where small angles of attack resulted in a higher critical surge speed, and thus, Reynolds number. For the smallest angle of attack of $3^{\circ}$, $Re_{c}$ was found to be 4000, whereas for $7^{\circ}$, $Re_{c}$ was 2600. Net circulation within the wake was accurately predicted by Wagner's function. The unsteady shear layer was characterised by its vorticity and streamwise velocity at various downstream locations from the starting vortex. Furthermore, the starting vortex was found to detach from the shear layer and attain about $33\%$ of the net circulation in the field in cases where secondary vortices do not form ($Re<Re_c$) and $\approx 42-47\%$ at higher surge speeds, where the shear layer breaks into Kelvin-Helmholtz type vortices ($Re>Re_c$). 


Despite detaching from the shear layer, the starting vortex was able to merge with neighbouring secondary vortices within its vicinity for $Re>Re_c$. For $Re<Re_c$, the starting vortex attains its max circulation at $t'_{F}= 0.16$ for all three angles of attack. For $Re>Re_c$, the corresponding $t'_{F}$ is between 0.21-0.25, after which the circulation drops and the starting vortex merges with a secondary vortex, finally reaching its final state at $1.5<t'_{\Gamma}<3.1$. If a limiting process exists for the starting vortex formed by an impulsively started airfoil, it was not observed for the present set of initial conditions. It is postulated that starting vortex detachment occurred as a result of competition between the velocity induced by the starting vortex towards itself and the opposite convective speed of the shear layer. This was ascertained qualitatively through forward and backward-time FTLEs, which highlighted regions of maximum particle separation and attraction, respectively. At time scales leading to detachment, the pFTLE showed local maxima between the starting vortex and the rest of the shear layer, indicating the separation of fluid particles. Once vortex detachment occurred, two distinct pFTLE ridges were obtained, where one is associated with the starting vortex and the other with the shear layer. Forward-time FTLE ridges highlighted the sensitivity of FTLEs to asymmetry, where parts of the starting vortex closer to the shear layer, which is associated with high velocity gradients, showed higher pFTLE values than the rest of the ridge enclosing the starting vortex. It was also observed that pFTLE ridges did not fully enclose the starting vortex at $t\sim t_{\mathrm{detachment}}$ as well as $t> t_{\mathrm{detachment}}$. Finally, vortex detachment was quantified by using a non-dimensional parameter, $\Gamma'$, originally devised by ~\citet{sattari2012growth}, which compares the velocity induced by the starting vortex on the shear layer to the convective speed of the shear layer, away from the starting vortex. The parameter was found to converge to a value of 1.5 at detachment, in agreement with ~\citet{sattari2012growth}'s observations. Through this study, it has been shown that a limiting process for vortex circulation entails detachment, however, vortex detachment does not always manifest due to a limiting process. Furthermore, the formation of secondary vortices does not necessarily occur as a result of starting vortex detachment. Thus, asymmetric vortices, formed without any rotational symmetry, exhibit more complex short-term dynamics than their axisymmetric equivalents. 
\nocite{*}
\bibliography{aipsamp}

\providecommand{\noopsort}[1]{}\providecommand{\singleletter}[1]{#1}%
\begin{thebibliography}{48}%
\makeatletter
\providecommand \@ifxundefined [1]{%
 \@ifx{#1\undefined}
}%
\providecommand \@ifnum [1]{%
 \ifnum #1\expandafter \@firstoftwo
 \else \expandafter \@secondoftwo
 \fi
}%
\providecommand \@ifx [1]{%
 \ifx #1\expandafter \@firstoftwo
 \else \expandafter \@secondoftwo
 \fi
}%
\providecommand \natexlab [1]{#1}%
\providecommand \enquote  [1]{``#1''}%
\providecommand \bibnamefont  [1]{#1}%
\providecommand \bibfnamefont [1]{#1}%
\providecommand \citenamefont [1]{#1}%
\providecommand \href@noop [0]{\@secondoftwo}%
\providecommand \href [0]{\begingroup \@sanitize@url \@href}%
\providecommand \@href[1]{\@@startlink{#1}\@@href}%
\providecommand \@@href[1]{\endgroup#1\@@endlink}%
\providecommand \@sanitize@url [0]{\catcode `\\12\catcode `\$12\catcode `\&12\catcode `\#12\catcode `\^12\catcode `\_12\catcode `\%12\relax}%
\providecommand \@@startlink[1]{}%
\providecommand \@@endlink[0]{}%
\providecommand \url  [0]{\begingroup\@sanitize@url \@url }%
\providecommand \@url [1]{\endgroup\@href {#1}{\urlprefix }}%
\providecommand \urlprefix  [0]{URL }%
\providecommand \Eprint [0]{\href }%
\providecommand \doibase [0]{http://dx.doi.org/}%
\providecommand \selectlanguage [0]{\@gobble}%
\providecommand \bibinfo  [0]{\@secondoftwo}%
\providecommand \bibfield  [0]{\@secondoftwo}%
\providecommand \translation [1]{[#1]}%
\providecommand \BibitemOpen [0]{}%
\providecommand \bibitemStop [0]{}%
\providecommand \bibitemNoStop [0]{.\EOS\space}%
\providecommand \EOS [0]{\spacefactor3000\relax}%
\providecommand \BibitemShut  [1]{\csname bibitem#1\endcsname}%
\let\auto@bib@innerbib\@empty
\bibitem [{\citenamefont {Dabiri}\ \emph {et~al.}(2005)\citenamefont {Dabiri}, \citenamefont {Colin}, \citenamefont {Costello},\ and\ \citenamefont {Gharib}}]{dabirijellyfish}%
  \BibitemOpen
  \bibfield  {author} {\bibinfo {author} {\bibfnamefont {J.~O.}\ \bibnamefont {Dabiri}}, \bibinfo {author} {\bibfnamefont {S.~P.}\ \bibnamefont {Colin}}, \bibinfo {author} {\bibfnamefont {J.~H.}\ \bibnamefont {Costello}}, \ and\ \bibinfo {author} {\bibfnamefont {M.}~\bibnamefont {Gharib}},\ }\bibfield  {title} {\enquote {\bibinfo {title} {Flow patterns generated by oblate medusan jellyfish: field measurements and laboratory analyses},}\ }\href@noop {} {\bibfield  {journal} {\bibinfo  {journal} {Journal of Experimental Biology}\ }\textbf {\bibinfo {volume} {208}},\ \bibinfo {pages} {1257--1265} (\bibinfo {year} {2005})}\BibitemShut {NoStop}%
\bibitem [{\citenamefont {Eldredge}\ and\ \citenamefont {Jones}(2019)}]{eldredge2019leading}%
  \BibitemOpen
  \bibfield  {author} {\bibinfo {author} {\bibfnamefont {J.~D.}\ \bibnamefont {Eldredge}}\ and\ \bibinfo {author} {\bibfnamefont {A.~R.}\ \bibnamefont {Jones}},\ }\bibfield  {title} {\enquote {\bibinfo {title} {Leading-edge vortices: mechanics and modeling},}\ }\href@noop {} {\bibfield  {journal} {\bibinfo  {journal} {Annual Review of Fluid Mechanics}\ }\textbf {\bibinfo {volume} {51}},\ \bibinfo {pages} {75--104} (\bibinfo {year} {2019})}\BibitemShut {NoStop}%
\bibitem [{\citenamefont {Cummins}\ \emph {et~al.}(2018)\citenamefont {Cummins}, \citenamefont {Seale}, \citenamefont {Macente}, \citenamefont {Certini}, \citenamefont {Mastropaolo}, \citenamefont {Viola},\ and\ \citenamefont {Nakayama}}]{dandy}%
  \BibitemOpen
  \bibfield  {author} {\bibinfo {author} {\bibfnamefont {C.}~\bibnamefont {Cummins}}, \bibinfo {author} {\bibfnamefont {M.}~\bibnamefont {Seale}}, \bibinfo {author} {\bibfnamefont {A.}~\bibnamefont {Macente}}, \bibinfo {author} {\bibfnamefont {D.}~\bibnamefont {Certini}}, \bibinfo {author} {\bibfnamefont {E.}~\bibnamefont {Mastropaolo}}, \bibinfo {author} {\bibfnamefont {I.~M.}\ \bibnamefont {Viola}}, \ and\ \bibinfo {author} {\bibfnamefont {N.}~\bibnamefont {Nakayama}},\ }\bibfield  {title} {\enquote {\bibinfo {title} {A separated vortex ring underlies the flight of the dandelion},}\ }\href@noop {} {\bibfield  {journal} {\bibinfo  {journal} {Nature}\ }\textbf {\bibinfo {volume} {562}},\ \bibinfo {pages} {414--418} (\bibinfo {year} {2018})}\BibitemShut {NoStop}%
\bibitem [{\citenamefont {Rival}, \citenamefont {Prangemeier},\ and\ \citenamefont {Tropea}(2010)}]{rival2010influence}%
  \BibitemOpen
  \bibfield  {author} {\bibinfo {author} {\bibfnamefont {D.}~\bibnamefont {Rival}}, \bibinfo {author} {\bibfnamefont {T.}~\bibnamefont {Prangemeier}}, \ and\ \bibinfo {author} {\bibfnamefont {C.}~\bibnamefont {Tropea}},\ }\bibfield  {title} {\enquote {\bibinfo {title} {The influence of airfoil kinematics on the formation of leading-edge vortices in bio-inspired flight},}\ }\href@noop {} {\bibfield  {journal} {\bibinfo  {journal} {Animal Locomotion}\ ,\ \bibinfo {pages} {261--271}} (\bibinfo {year} {2010})}\BibitemShut {NoStop}%
\bibitem [{\citenamefont {Andreu-Angulo}\ \emph {et~al.}(2020)\citenamefont {Andreu-Angulo}, \citenamefont {Babinsky}, \citenamefont {Biler}, \citenamefont {Sedky},\ and\ \citenamefont {Jones}}]{gust}%
  \BibitemOpen
  \bibfield  {author} {\bibinfo {author} {\bibfnamefont {I.}~\bibnamefont {Andreu-Angulo}}, \bibinfo {author} {\bibfnamefont {H.}~\bibnamefont {Babinsky}}, \bibinfo {author} {\bibfnamefont {H.}~\bibnamefont {Biler}}, \bibinfo {author} {\bibfnamefont {G.}~\bibnamefont {Sedky}}, \ and\ \bibinfo {author} {\bibfnamefont {A.~R.}\ \bibnamefont {Jones}},\ }\bibfield  {title} {\enquote {\bibinfo {title} {Effect of transverse gust velocity profiles},}\ }\href@noop {} {\bibfield  {journal} {\bibinfo  {journal} {AIAA Journal}\ }\textbf {\bibinfo {volume} {58}},\ \bibinfo {pages} {5123--5133} (\bibinfo {year} {2020})}\BibitemShut {NoStop}%
\bibitem [{\citenamefont {Sequeira}\ and\ \citenamefont {Miller}(2014)}]{tidal_turbines}%
  \BibitemOpen
  \bibfield  {author} {\bibinfo {author} {\bibfnamefont {C.~L.}\ \bibnamefont {Sequeira}}\ and\ \bibinfo {author} {\bibfnamefont {R.~J.}\ \bibnamefont {Miller}},\ }\bibfield  {title} {\enquote {\bibinfo {title} {Unsteady gust response of tidal stream turbines},}\ }in\ \href {\doibase 10.1109/OCEANS.2014.7003026} {\emph {\bibinfo {booktitle} {2014 Oceans - St. John's}}}\ (\bibinfo {year} {2014})\ pp.\ \bibinfo {pages} {1--10}\BibitemShut {NoStop}%
\bibitem [{\citenamefont {Gharib}, \citenamefont {Rambod},\ and\ \citenamefont {Shariff}(1998)}]{gharib1998universal}%
  \BibitemOpen
  \bibfield  {author} {\bibinfo {author} {\bibfnamefont {M.}~\bibnamefont {Gharib}}, \bibinfo {author} {\bibfnamefont {E.}~\bibnamefont {Rambod}}, \ and\ \bibinfo {author} {\bibfnamefont {K.}~\bibnamefont {Shariff}},\ }\bibfield  {title} {\enquote {\bibinfo {title} {A universal time scale for vortex ring formation},}\ }\href@noop {} {\bibfield  {journal} {\bibinfo  {journal} {Journal of Fluid Mechanics}\ }\textbf {\bibinfo {volume} {360}},\ \bibinfo {pages} {121--140} (\bibinfo {year} {1998})}\BibitemShut {NoStop}%
\bibitem [{\citenamefont {Didden}(1979)}]{didden1979formation}%
  \BibitemOpen
  \bibfield  {author} {\bibinfo {author} {\bibfnamefont {N.}~\bibnamefont {Didden}},\ }\bibfield  {title} {\enquote {\bibinfo {title} {On the formation of vortex rings: rolling-up and production of circulation},}\ }\href@noop {} {\bibfield  {journal} {\bibinfo  {journal} {Zeitschrift f{\"u}r angewandte Mathematik und Physik ZAMP}\ }\textbf {\bibinfo {volume} {30}},\ \bibinfo {pages} {101--116} (\bibinfo {year} {1979})}\BibitemShut {NoStop}%
\bibitem [{\citenamefont {Jeon}\ and\ \citenamefont {Gharib}(2004)}]{jeon2004relationship}%
  \BibitemOpen
  \bibfield  {author} {\bibinfo {author} {\bibfnamefont {D.}~\bibnamefont {Jeon}}\ and\ \bibinfo {author} {\bibfnamefont {M.}~\bibnamefont {Gharib}},\ }\bibfield  {title} {\enquote {\bibinfo {title} {On the relationship between the vortex formation process and cylinder wake vortex patterns},}\ }\href@noop {} {\bibfield  {journal} {\bibinfo  {journal} {Journal of Fluid Mechanics}\ }\textbf {\bibinfo {volume} {519}},\ \bibinfo {pages} {161--181} (\bibinfo {year} {2004})}\BibitemShut {NoStop}%
\bibitem [{\citenamefont {Ringuette}, \citenamefont {Milano},\ and\ \citenamefont {Gharib}(2007)}]{ringuette2007role}%
  \BibitemOpen
  \bibfield  {author} {\bibinfo {author} {\bibfnamefont {M.~J.}\ \bibnamefont {Ringuette}}, \bibinfo {author} {\bibfnamefont {M.}~\bibnamefont {Milano}}, \ and\ \bibinfo {author} {\bibfnamefont {M.}~\bibnamefont {Gharib}},\ }\bibfield  {title} {\enquote {\bibinfo {title} {Role of the tip vortex in the force generation of low-aspect-ratio normal flat plates},}\ }\href@noop {} {\bibfield  {journal} {\bibinfo  {journal} {Journal of Fluid Mechanics}\ }\textbf {\bibinfo {volume} {581}},\ \bibinfo {pages} {453--468} (\bibinfo {year} {2007})}\BibitemShut {NoStop}%
\bibitem [{\citenamefont {Milano}\ and\ \citenamefont {Gharib}(2005)}]{milano_gharib}%
  \BibitemOpen
  \bibfield  {author} {\bibinfo {author} {\bibfnamefont {M.}~\bibnamefont {Milano}}\ and\ \bibinfo {author} {\bibfnamefont {M.}~\bibnamefont {Gharib}},\ }\bibfield  {title} {\enquote {\bibinfo {title} {Uncovering the physics of flapping flat plates with artificial evolution},}\ }\href@noop {} {\bibfield  {journal} {\bibinfo  {journal} {Journal of Fluid Mechanics}\ }\textbf {\bibinfo {volume} {534}},\ \bibinfo {pages} {403--409} (\bibinfo {year} {2005})}\BibitemShut {NoStop}%
\bibitem [{\citenamefont {Dabiri}\ and\ \citenamefont {Gharib}(2005)}]{dabiri2005starting}%
  \BibitemOpen
  \bibfield  {author} {\bibinfo {author} {\bibfnamefont {J.~O.}\ \bibnamefont {Dabiri}}\ and\ \bibinfo {author} {\bibfnamefont {M.}~\bibnamefont {Gharib}},\ }\bibfield  {title} {\enquote {\bibinfo {title} {Starting flow through nozzles with temporally variable exit diameter},}\ }\href@noop {} {\bibfield  {journal} {\bibinfo  {journal} {Journal of Fluid Mechanics}\ }\textbf {\bibinfo {volume} {538}},\ \bibinfo {pages} {111--136} (\bibinfo {year} {2005})}\BibitemShut {NoStop}%
\bibitem [{\citenamefont {Thomson}(1883)}]{thomson1883treatise}%
  \BibitemOpen
  \bibfield  {author} {\bibinfo {author} {\bibfnamefont {J.~J.}\ \bibnamefont {Thomson}},\ }\href@noop {} {\emph {\bibinfo {title} {A Treatise on the Motion of Vortex Rings: an essay to which the Adams prize was adjudged in 1882, in the University of Cambridge}}}\ (\bibinfo  {publisher} {Macmillan},\ \bibinfo {year} {1883})\BibitemShut {NoStop}%
\bibitem [{\citenamefont {Benjamin}(1976)}]{benjamin1976alliance}%
  \BibitemOpen
  \bibfield  {author} {\bibinfo {author} {\bibfnamefont {T.~B.}\ \bibnamefont {Benjamin}},\ }\bibfield  {title} {\enquote {\bibinfo {title} {The alliance of practical and analytical insights into the nonlinear problems of fluid mechanics},}\ }in\ \href@noop {} {\emph {\bibinfo {booktitle} {Applications of Methods of Functional Analysis to Problems in Mechanics: Joint Symposium IUTAM/IMU Held in Marseille, September 1--6, 1975}}}\ (\bibinfo {organization} {Springer},\ \bibinfo {year} {1976})\ pp.\ \bibinfo {pages} {8--29}\BibitemShut {NoStop}%
\bibitem [{\citenamefont {Shusser}\ and\ \citenamefont {Gharib}(2000)}]{shusser2000energy}%
  \BibitemOpen
  \bibfield  {author} {\bibinfo {author} {\bibfnamefont {M.}~\bibnamefont {Shusser}}\ and\ \bibinfo {author} {\bibfnamefont {M.}~\bibnamefont {Gharib}},\ }\bibfield  {title} {\enquote {\bibinfo {title} {Energy and velocity of a forming vortex ring},}\ }\href@noop {} {\bibfield  {journal} {\bibinfo  {journal} {Physics of Fluids}\ }\textbf {\bibinfo {volume} {12}},\ \bibinfo {pages} {618--621} (\bibinfo {year} {2000})}\BibitemShut {NoStop}%
\bibitem [{\citenamefont {Limbourg}\ and\ \citenamefont {Nedi{\'c}}(2021{\natexlab{a}})}]{limbourg2021asymptotic}%
  \BibitemOpen
  \bibfield  {author} {\bibinfo {author} {\bibfnamefont {R.}~\bibnamefont {Limbourg}}\ and\ \bibinfo {author} {\bibfnamefont {J.}~\bibnamefont {Nedi{\'c}}},\ }\bibfield  {title} {\enquote {\bibinfo {title} {On the asymptotic matching procedure predicting the formation number},}\ }\href@noop {} {\bibfield  {journal} {\bibinfo  {journal} {Physics of Fluids}\ }\textbf {\bibinfo {volume} {33}} (\bibinfo {year} {2021}{\natexlab{a}})}\BibitemShut {NoStop}%
\bibitem [{\citenamefont {DeVoria}\ and\ \citenamefont {Ringuette}(2012)}]{devorianringuette}%
  \BibitemOpen
  \bibfield  {author} {\bibinfo {author} {\bibfnamefont {A.~C.}\ \bibnamefont {DeVoria}}\ and\ \bibinfo {author} {\bibfnamefont {M.~J.}\ \bibnamefont {Ringuette}},\ }\bibfield  {title} {\enquote {\bibinfo {title} {Vortex formation and saturation for low-aspect-ratio rotating flat-plate fins},}\ }\href@noop {} {\bibfield  {journal} {\bibinfo  {journal} {Experiments in fluids}\ }\textbf {\bibinfo {volume} {52}},\ \bibinfo {pages} {441--462} (\bibinfo {year} {2012})}\BibitemShut {NoStop}%
\bibitem [{\citenamefont {Limbourg}\ and\ \citenamefont {Nedi{\'c}}(2021{\natexlab{b}})}]{limbourg2021formation}%
  \BibitemOpen
  \bibfield  {author} {\bibinfo {author} {\bibfnamefont {R.}~\bibnamefont {Limbourg}}\ and\ \bibinfo {author} {\bibfnamefont {J.}~\bibnamefont {Nedi{\'c}}},\ }\bibfield  {title} {\enquote {\bibinfo {title} {Formation of an orifice-generated vortex ring},}\ }\href@noop {} {\bibfield  {journal} {\bibinfo  {journal} {Journal of Fluid Mechanics}\ }\textbf {\bibinfo {volume} {913}},\ \bibinfo {pages} {A29} (\bibinfo {year} {2021}{\natexlab{b}})}\BibitemShut {NoStop}%
\bibitem [{\citenamefont {Dabiri}(2009)}]{dabiri2009optimal}%
  \BibitemOpen
  \bibfield  {author} {\bibinfo {author} {\bibfnamefont {J.~O.}\ \bibnamefont {Dabiri}},\ }\bibfield  {title} {\enquote {\bibinfo {title} {Optimal vortex formation as a unifying principle in biological propulsion},}\ }\href@noop {} {\bibfield  {journal} {\bibinfo  {journal} {Annual review of fluid mechanics}\ }\textbf {\bibinfo {volume} {41}},\ \bibinfo {pages} {17--33} (\bibinfo {year} {2009})}\BibitemShut {NoStop}%
\bibitem [{\citenamefont {Limbourg}\ and\ \citenamefont {Nedi{\'c}}(2021{\natexlab{c}})}]{limbourg2021extension}%
  \BibitemOpen
  \bibfield  {author} {\bibinfo {author} {\bibfnamefont {R.}~\bibnamefont {Limbourg}}\ and\ \bibinfo {author} {\bibfnamefont {J.}~\bibnamefont {Nedi{\'c}}},\ }\bibfield  {title} {\enquote {\bibinfo {title} {An extension to the universal time scale for vortex ring formation},}\ }\href@noop {} {\bibfield  {journal} {\bibinfo  {journal} {Journal of Fluid Mechanics}\ }\textbf {\bibinfo {volume} {915}},\ \bibinfo {pages} {A46} (\bibinfo {year} {2021}{\natexlab{c}})}\BibitemShut {NoStop}%
\bibitem [{\citenamefont {Limbourg}\ and\ \citenamefont {Nedi{\'c}}(2021{\natexlab{d}})}]{limbourg2021extended}%
  \BibitemOpen
  \bibfield  {author} {\bibinfo {author} {\bibfnamefont {R.}~\bibnamefont {Limbourg}}\ and\ \bibinfo {author} {\bibfnamefont {J.}~\bibnamefont {Nedi{\'c}}},\ }\bibfield  {title} {\enquote {\bibinfo {title} {An extended model for orifice starting jets},}\ }\href@noop {} {\bibfield  {journal} {\bibinfo  {journal} {Physics of Fluids}\ }\textbf {\bibinfo {volume} {33}} (\bibinfo {year} {2021}{\natexlab{d}})}\BibitemShut {NoStop}%
\bibitem [{\citenamefont {O’Farrell}\ and\ \citenamefont {Dabiri}(2014)}]{o2014pinch}%
  \BibitemOpen
  \bibfield  {author} {\bibinfo {author} {\bibfnamefont {C.}~\bibnamefont {O’Farrell}}\ and\ \bibinfo {author} {\bibfnamefont {J.~O.}\ \bibnamefont {Dabiri}},\ }\bibfield  {title} {\enquote {\bibinfo {title} {Pinch-off of non-axisymmetric vortex rings},}\ }\href@noop {} {\bibfield  {journal} {\bibinfo  {journal} {Journal of fluid mechanics}\ }\textbf {\bibinfo {volume} {740}},\ \bibinfo {pages} {61--96} (\bibinfo {year} {2014})}\BibitemShut {NoStop}%
\bibitem [{\citenamefont {Prandtl}(1924)}]{prandtl}%
  \BibitemOpen
  \bibfield  {author} {\bibinfo {author} {\bibfnamefont {L.}~\bibnamefont {Prandtl}},\ }\bibfield  {title} {\enquote {\bibinfo {title} {{\"U}ber die entstehung von wirbeln in der idealen fl{\"u}ssigkeit, mit anwendung auf die tragfl{\"u}geltheorie und andere aufgaben},}\ }\href@noop {} {\bibfield  {journal} {\bibinfo  {journal} {Vortr{\"a}ge aus dem Gebiete der Hydro-und Aerodynamik (Innsbruck 1922)}\ ,\ \bibinfo {pages} {18--33}} (\bibinfo {year} {1924})}\BibitemShut {NoStop}%
\bibitem [{\citenamefont {Pullin}\ and\ \citenamefont {Perry}(1980)}]{perry}%
  \BibitemOpen
  \bibfield  {author} {\bibinfo {author} {\bibfnamefont {D.}~\bibnamefont {Pullin}}\ and\ \bibinfo {author} {\bibfnamefont {A.}~\bibnamefont {Perry}},\ }\bibfield  {title} {\enquote {\bibinfo {title} {Some flow visualization experiments on the starting vortex},}\ }\href@noop {} {\bibfield  {journal} {\bibinfo  {journal} {Journal of Fluid Mechanics}\ }\textbf {\bibinfo {volume} {97}},\ \bibinfo {pages} {239--255} (\bibinfo {year} {1980})}\BibitemShut {NoStop}%
\bibitem [{\citenamefont {Auerbach}(1987)}]{auerbach}%
  \BibitemOpen
  \bibfield  {author} {\bibinfo {author} {\bibfnamefont {D.}~\bibnamefont {Auerbach}},\ }\bibfield  {title} {\enquote {\bibinfo {title} {Experiments on the trajectory and circulation of the starting vortex},}\ }\href@noop {} {\bibfield  {journal} {\bibinfo  {journal} {Journal of Fluid Mechanics}\ }\textbf {\bibinfo {volume} {183}},\ \bibinfo {pages} {185--198} (\bibinfo {year} {1987})}\BibitemShut {NoStop}%
\bibitem [{\citenamefont {Whalley}\ and\ \citenamefont {Choi}(2012)}]{whalley_starting}%
  \BibitemOpen
  \bibfield  {author} {\bibinfo {author} {\bibfnamefont {R.~D.}\ \bibnamefont {Whalley}}\ and\ \bibinfo {author} {\bibfnamefont {K.-S.}\ \bibnamefont {Choi}},\ }\bibfield  {title} {\enquote {\bibinfo {title} {The starting vortex in quiescent air induced by dielectric-barrier-discharge plasma},}\ }\href@noop {} {\bibfield  {journal} {\bibinfo  {journal} {Journal of Fluid Mechanics}\ }\textbf {\bibinfo {volume} {703}},\ \bibinfo {pages} {192--203} (\bibinfo {year} {2012})}\BibitemShut {NoStop}%
\bibitem [{\citenamefont {Luchini}\ and\ \citenamefont {Tognaccini}(2002)}]{luchini2002start}%
  \BibitemOpen
  \bibfield  {author} {\bibinfo {author} {\bibfnamefont {P.}~\bibnamefont {Luchini}}\ and\ \bibinfo {author} {\bibfnamefont {R.}~\bibnamefont {Tognaccini}},\ }\bibfield  {title} {\enquote {\bibinfo {title} {The start-up vortex issuing from a semi-infinite flat plate},}\ }\href@noop {} {\bibfield  {journal} {\bibinfo  {journal} {Journal of Fluid Mechanics}\ }\textbf {\bibinfo {volume} {455}},\ \bibinfo {pages} {175--193} (\bibinfo {year} {2002})}\BibitemShut {NoStop}%
\bibitem [{\citenamefont {Xu}, \citenamefont {Nitsche},\ and\ \citenamefont {Krasny}(2017)}]{xu2017computation}%
  \BibitemOpen
  \bibfield  {author} {\bibinfo {author} {\bibfnamefont {L.}~\bibnamefont {Xu}}, \bibinfo {author} {\bibfnamefont {M.}~\bibnamefont {Nitsche}}, \ and\ \bibinfo {author} {\bibfnamefont {R.}~\bibnamefont {Krasny}},\ }\bibfield  {title} {\enquote {\bibinfo {title} {Computation of the starting vortex flow past a flat plate},}\ }\href@noop {} {\bibfield  {journal} {\bibinfo  {journal} {Procedia IUTAM}\ }\textbf {\bibinfo {volume} {20}},\ \bibinfo {pages} {136--143} (\bibinfo {year} {2017})}\BibitemShut {NoStop}%
\bibitem [{\citenamefont {Kaden}(1931)}]{kaden}%
  \BibitemOpen
  \bibfield  {author} {\bibinfo {author} {\bibfnamefont {H.}~\bibnamefont {Kaden}},\ }\bibfield  {title} {\enquote {\bibinfo {title} {Aufwicklung einer unstabilen unstetigkeitsfl{\"a}che},}\ }\href@noop {} {\bibfield  {journal} {\bibinfo  {journal} {Ingenieur-Archiv}\ }\textbf {\bibinfo {volume} {2}},\ \bibinfo {pages} {140--168} (\bibinfo {year} {1931})}\BibitemShut {NoStop}%
\bibitem [{\citenamefont {Pullin}(1978)}]{pullin1}%
  \BibitemOpen
  \bibfield  {author} {\bibinfo {author} {\bibfnamefont {D.}~\bibnamefont {Pullin}},\ }\bibfield  {title} {\enquote {\bibinfo {title} {The large-scale structure of unsteady self-similar rolled-up vortex sheets},}\ }\href@noop {} {\bibfield  {journal} {\bibinfo  {journal} {Journal of Fluid Mechanics}\ }\textbf {\bibinfo {volume} {88}},\ \bibinfo {pages} {401--430} (\bibinfo {year} {1978})}\BibitemShut {NoStop}%
\bibitem [{\citenamefont {Pullin}\ and\ \citenamefont {Sader}(2021)}]{pullin2021}%
  \BibitemOpen
  \bibfield  {author} {\bibinfo {author} {\bibfnamefont {D.}~\bibnamefont {Pullin}}\ and\ \bibinfo {author} {\bibfnamefont {J.~E.}\ \bibnamefont {Sader}},\ }\bibfield  {title} {\enquote {\bibinfo {title} {On the starting vortex generated by a translating and rotating flat plate},}\ }\href@noop {} {\bibfield  {journal} {\bibinfo  {journal} {Journal of Fluid Mechanics}\ }\textbf {\bibinfo {volume} {906}},\ \bibinfo {pages} {A9} (\bibinfo {year} {2021})}\BibitemShut {NoStop}%
\bibitem [{\citenamefont {Hinton}\ \emph {et~al.}(2024)\citenamefont {Hinton}, \citenamefont {Leonard}, \citenamefont {Pullin},\ and\ \citenamefont {Sader}}]{pullin_arbit}%
  \BibitemOpen
  \bibfield  {author} {\bibinfo {author} {\bibfnamefont {E.~M.}\ \bibnamefont {Hinton}}, \bibinfo {author} {\bibfnamefont {A.}~\bibnamefont {Leonard}}, \bibinfo {author} {\bibfnamefont {D.}~\bibnamefont {Pullin}}, \ and\ \bibinfo {author} {\bibfnamefont {J.~E.}\ \bibnamefont {Sader}},\ }\bibfield  {title} {\enquote {\bibinfo {title} {Starting vortices generated by an arbitrary solid body with any number of edges},}\ }\href@noop {} {\bibfield  {journal} {\bibinfo  {journal} {Journal of Fluid Mechanics}\ }\textbf {\bibinfo {volume} {987}},\ \bibinfo {pages} {A11} (\bibinfo {year} {2024})}\BibitemShut {NoStop}%
\bibitem [{\citenamefont {Leweke}\ and\ \citenamefont {Williamson}(1998)}]{leweke1998cooperative}%
  \BibitemOpen
  \bibfield  {author} {\bibinfo {author} {\bibfnamefont {T.}~\bibnamefont {Leweke}}\ and\ \bibinfo {author} {\bibfnamefont {C.~H.}\ \bibnamefont {Williamson}},\ }\bibfield  {title} {\enquote {\bibinfo {title} {Cooperative elliptic instability of a vortex pair},}\ }\href@noop {} {\bibfield  {journal} {\bibinfo  {journal} {Journal of fluid mechanics}\ }\textbf {\bibinfo {volume} {360}},\ \bibinfo {pages} {85--119} (\bibinfo {year} {1998})}\BibitemShut {NoStop}%
\bibitem [{\citenamefont {Walker}\ \emph {et~al.}(1987)\citenamefont {Walker}, \citenamefont {Smith}, \citenamefont {Cerra},\ and\ \citenamefont {Doligalski}}]{walker1987impact}%
  \BibitemOpen
  \bibfield  {author} {\bibinfo {author} {\bibfnamefont {J.}~\bibnamefont {Walker}}, \bibinfo {author} {\bibfnamefont {C.}~\bibnamefont {Smith}}, \bibinfo {author} {\bibfnamefont {A.}~\bibnamefont {Cerra}}, \ and\ \bibinfo {author} {\bibfnamefont {T.}~\bibnamefont {Doligalski}},\ }\bibfield  {title} {\enquote {\bibinfo {title} {The impact of a vortex ring on a wall},}\ }\href@noop {} {\bibfield  {journal} {\bibinfo  {journal} {Journal of Fluid Mechanics}\ }\textbf {\bibinfo {volume} {181}},\ \bibinfo {pages} {99--140} (\bibinfo {year} {1987})}\BibitemShut {NoStop}%
\bibitem [{\citenamefont {Huang}\ \emph {et~al.}(2001)\citenamefont {Huang}, \citenamefont {Wu}, \citenamefont {Jeng},\ and\ \citenamefont {Chen}}]{huang_wu_jeng_chen_2001}%
  \BibitemOpen
  \bibfield  {author} {\bibinfo {author} {\bibfnamefont {R.~F.}\ \bibnamefont {Huang}}, \bibinfo {author} {\bibfnamefont {J.~Y.}\ \bibnamefont {Wu}}, \bibinfo {author} {\bibfnamefont {J.~H.}\ \bibnamefont {Jeng}}, \ and\ \bibinfo {author} {\bibfnamefont {R.~C.}\ \bibnamefont {Chen}},\ }\bibfield  {title} {\enquote {\bibinfo {title} {Surface flow and vortex shedding of an impulsively started wing},}\ }\href {\doibase 10.1017/S002211200100489X} {\bibfield  {journal} {\bibinfo  {journal} {Journal of Fluid Mechanics}\ }\textbf {\bibinfo {volume} {441}},\ \bibinfo {pages} {265–292} (\bibinfo {year} {2001})}\BibitemShut {NoStop}%
\bibitem [{\citenamefont {Wagner}(1925)}]{wagner}%
  \BibitemOpen
  \bibfield  {author} {\bibinfo {author} {\bibfnamefont {H.}~\bibnamefont {Wagner}},\ }\bibfield  {title} {\enquote {\bibinfo {title} {Über die entstehung des dynamischen auftriebes von tragflügeln},}\ }\href@noop {} {\bibfield  {journal} {\bibinfo  {journal} {ZAMM - Journal of Applied Mathematics and Mechanics / Zeitschrift für Angewandte Mathematik und Mechanik}\ }\textbf {\bibinfo {volume} {5}},\ \bibinfo {pages} {17--35} (\bibinfo {year} {1925})}\BibitemShut {NoStop}%
\bibitem [{\citenamefont {Beckwith}\ and\ \citenamefont {Babinsky}(2009)}]{beckwith2009impulsively}%
  \BibitemOpen
  \bibfield  {author} {\bibinfo {author} {\bibfnamefont {R.}~\bibnamefont {Beckwith}}\ and\ \bibinfo {author} {\bibfnamefont {H.}~\bibnamefont {Babinsky}},\ }\bibfield  {title} {\enquote {\bibinfo {title} {Impulsively started flat plate flow},}\ }\href@noop {} {\bibfield  {journal} {\bibinfo  {journal} {Journal of aircraft}\ }\textbf {\bibinfo {volume} {46}},\ \bibinfo {pages} {2186--2189} (\bibinfo {year} {2009})}\BibitemShut {NoStop}%
\bibitem [{\citenamefont {Baskaran}\ and\ \citenamefont {Mulleners}(2022)}]{baskaran2022lagrangian}%
  \BibitemOpen
  \bibfield  {author} {\bibinfo {author} {\bibfnamefont {M.}~\bibnamefont {Baskaran}}\ and\ \bibinfo {author} {\bibfnamefont {K.}~\bibnamefont {Mulleners}},\ }\bibfield  {title} {\enquote {\bibinfo {title} {Lagrangian analysis of bio-inspired vortex ring formation},}\ }\href@noop {} {\bibfield  {journal} {\bibinfo  {journal} {Flow}\ }\textbf {\bibinfo {volume} {2}},\ \bibinfo {pages} {E16} (\bibinfo {year} {2022})}\BibitemShut {NoStop}%
\bibitem [{\citenamefont {Francescangeli}\ and\ \citenamefont {Mulleners}(2021)}]{francescangeli2021discrete}%
  \BibitemOpen
  \bibfield  {author} {\bibinfo {author} {\bibfnamefont {D.}~\bibnamefont {Francescangeli}}\ and\ \bibinfo {author} {\bibfnamefont {K.}~\bibnamefont {Mulleners}},\ }\bibfield  {title} {\enquote {\bibinfo {title} {Discrete shedding of secondary vortices along a modified kaden spiral},}\ }\href@noop {} {\bibfield  {journal} {\bibinfo  {journal} {Journal of Fluid Mechanics}\ }\textbf {\bibinfo {volume} {917}},\ \bibinfo {pages} {A44} (\bibinfo {year} {2021})}\BibitemShut {NoStop}%
\bibitem [{\citenamefont {de~Guyon}\ and\ \citenamefont {Mulleners}(2021)}]{karen_conical}%
  \BibitemOpen
  \bibfield  {author} {\bibinfo {author} {\bibfnamefont {G.}~\bibnamefont {de~Guyon}}\ and\ \bibinfo {author} {\bibfnamefont {K.}~\bibnamefont {Mulleners}},\ }\bibfield  {title} {\enquote {\bibinfo {title} {Scaling of the translational velocity of vortex rings behind conical objects},}\ }\href@noop {} {\bibfield  {journal} {\bibinfo  {journal} {Physical Review Fluids}\ }\textbf {\bibinfo {volume} {6}},\ \bibinfo {pages} {024701} (\bibinfo {year} {2021})}\BibitemShut {NoStop}%
\bibitem [{\citenamefont {Gan}, \citenamefont {Dawson},\ and\ \citenamefont {Nickels}(2012)}]{gan2012drag}%
  \BibitemOpen
  \bibfield  {author} {\bibinfo {author} {\bibfnamefont {L.}~\bibnamefont {Gan}}, \bibinfo {author} {\bibfnamefont {J.}~\bibnamefont {Dawson}}, \ and\ \bibinfo {author} {\bibfnamefont {T.}~\bibnamefont {Nickels}},\ }\bibfield  {title} {\enquote {\bibinfo {title} {On the drag of turbulent vortex rings},}\ }\href@noop {} {\bibfield  {journal} {\bibinfo  {journal} {Journal of Fluid Mechanics}\ }\textbf {\bibinfo {volume} {709}},\ \bibinfo {pages} {85--105} (\bibinfo {year} {2012})}\BibitemShut {NoStop}%
\bibitem [{\citenamefont {Sattari}\ \emph {et~al.}(2012)\citenamefont {Sattari}, \citenamefont {Rival}, \citenamefont {Martinuzzi},\ and\ \citenamefont {Tropea}}]{sattari2012growth}%
  \BibitemOpen
  \bibfield  {author} {\bibinfo {author} {\bibfnamefont {P.}~\bibnamefont {Sattari}}, \bibinfo {author} {\bibfnamefont {D.~E.}\ \bibnamefont {Rival}}, \bibinfo {author} {\bibfnamefont {R.~J.}\ \bibnamefont {Martinuzzi}}, \ and\ \bibinfo {author} {\bibfnamefont {C.}~\bibnamefont {Tropea}},\ }\bibfield  {title} {\enquote {\bibinfo {title} {Growth and separation of a start-up vortex from a two-dimensional shear layer},}\ }\href@noop {} {\bibfield  {journal} {\bibinfo  {journal} {Physics of Fluids}\ }\textbf {\bibinfo {volume} {24}} (\bibinfo {year} {2012})}\BibitemShut {NoStop}%
\bibitem [{\citenamefont {Shadden}, \citenamefont {Dabiri},\ and\ \citenamefont {Marsden}(2006)}]{shadden2006lagrangian}%
  \BibitemOpen
  \bibfield  {author} {\bibinfo {author} {\bibfnamefont {S.~C.}\ \bibnamefont {Shadden}}, \bibinfo {author} {\bibfnamefont {J.~O.}\ \bibnamefont {Dabiri}}, \ and\ \bibinfo {author} {\bibfnamefont {J.~E.}\ \bibnamefont {Marsden}},\ }\bibfield  {title} {\enquote {\bibinfo {title} {Lagrangian analysis of fluid transport in empirical vortex ring flows},}\ }\href@noop {} {\bibfield  {journal} {\bibinfo  {journal} {Physics of fluids}\ }\textbf {\bibinfo {volume} {18}} (\bibinfo {year} {2006})}\BibitemShut {NoStop}%
\bibitem [{\citenamefont {Shadden}\ \emph {et~al.}(2007)\citenamefont {Shadden}, \citenamefont {Katija}, \citenamefont {Rosenfeld}, \citenamefont {Marsden},\ and\ \citenamefont {Dabiri}}]{shadden2007transport}%
  \BibitemOpen
  \bibfield  {author} {\bibinfo {author} {\bibfnamefont {S.~C.}\ \bibnamefont {Shadden}}, \bibinfo {author} {\bibfnamefont {K.}~\bibnamefont {Katija}}, \bibinfo {author} {\bibfnamefont {M.}~\bibnamefont {Rosenfeld}}, \bibinfo {author} {\bibfnamefont {J.~E.}\ \bibnamefont {Marsden}}, \ and\ \bibinfo {author} {\bibfnamefont {J.~O.}\ \bibnamefont {Dabiri}},\ }\bibfield  {title} {\enquote {\bibinfo {title} {Transport and stirring induced by vortex formation},}\ }\href@noop {} {\bibfield  {journal} {\bibinfo  {journal} {Journal of Fluid Mechanics}\ }\textbf {\bibinfo {volume} {593}},\ \bibinfo {pages} {315--331} (\bibinfo {year} {2007})}\BibitemShut {NoStop}%
\bibitem [{\citenamefont {Green}, \citenamefont {Rowley},\ and\ \citenamefont {Haller}(2007)}]{green2007detection}%
  \BibitemOpen
  \bibfield  {author} {\bibinfo {author} {\bibfnamefont {M.~A.}\ \bibnamefont {Green}}, \bibinfo {author} {\bibfnamefont {C.~W.}\ \bibnamefont {Rowley}}, \ and\ \bibinfo {author} {\bibfnamefont {G.}~\bibnamefont {Haller}},\ }\bibfield  {title} {\enquote {\bibinfo {title} {Detection of lagrangian coherent structures in three-dimensional turbulence},}\ }\href@noop {} {\bibfield  {journal} {\bibinfo  {journal} {Journal of Fluid Mechanics}\ }\textbf {\bibinfo {volume} {572}},\ \bibinfo {pages} {111--120} (\bibinfo {year} {2007})}\BibitemShut {NoStop}%
\bibitem [{\citenamefont {O’Farrell}\ and\ \citenamefont {Dabiri}(2010)}]{o2010lagrangian}%
  \BibitemOpen
  \bibfield  {author} {\bibinfo {author} {\bibfnamefont {C.}~\bibnamefont {O’Farrell}}\ and\ \bibinfo {author} {\bibfnamefont {J.~O.}\ \bibnamefont {Dabiri}},\ }\bibfield  {title} {\enquote {\bibinfo {title} {A lagrangian approach to identifying vortex pinch-off},}\ }\href@noop {} {\bibfield  {journal} {\bibinfo  {journal} {Chaos: An Interdisciplinary Journal of Nonlinear Science}\ }\textbf {\bibinfo {volume} {20}} (\bibinfo {year} {2010})}\BibitemShut {NoStop}%
\bibitem [{\citenamefont {Mulleners}\ and\ \citenamefont {Raffel}(2012)}]{mulleners2012onset}%
  \BibitemOpen
  \bibfield  {author} {\bibinfo {author} {\bibfnamefont {K.}~\bibnamefont {Mulleners}}\ and\ \bibinfo {author} {\bibfnamefont {M.}~\bibnamefont {Raffel}},\ }\bibfield  {title} {\enquote {\bibinfo {title} {The onset of dynamic stall revisited},}\ }\href@noop {} {\bibfield  {journal} {\bibinfo  {journal} {Experiments in fluids}\ }\textbf {\bibinfo {volume} {52}},\ \bibinfo {pages} {779--793} (\bibinfo {year} {2012})}\BibitemShut {NoStop}%
\bibitem [{\citenamefont {Krishna}, \citenamefont {Green},\ and\ \citenamefont {Mulleners}(2018)}]{krishna2018flowfield}%
  \BibitemOpen
  \bibfield  {author} {\bibinfo {author} {\bibfnamefont {S.}~\bibnamefont {Krishna}}, \bibinfo {author} {\bibfnamefont {M.~A.}\ \bibnamefont {Green}}, \ and\ \bibinfo {author} {\bibfnamefont {K.}~\bibnamefont {Mulleners}},\ }\bibfield  {title} {\enquote {\bibinfo {title} {Flowfield and force evolution for a symmetric hovering flat-plate wing},}\ }\href@noop {} {\bibfield  {journal} {\bibinfo  {journal} {AIAA journal}\ }\textbf {\bibinfo {volume} {56}},\ \bibinfo {pages} {1360--1371} (\bibinfo {year} {2018})}\BibitemShut {NoStop}%
\end{thebibliography}%

\end{document}